\documentclass[11pt]{article} 
\usepackage{mystyle-new}
\usepackage{epsfig,amsmath} 
\usepackage{hepnames,hepunits}
\usepackage{hyperref}
\usepackage{color}
\usepackage{graphicx}
\definecolor{red}{rgb}{1,0,0}
\def\lesssim{\ \hbox{\raise 2pt \hbox{$<$} \kern -13pt
                     \lower 3pt \hbox{$\sim$}}\ }
\def\greatersim{\ \hbox{\raise 2pt \hbox{$>$} \kern -13pt
                     \lower 3pt \hbox{$\sim$}}\ }

\def\cascade{{\sc Cascade}}
\def\pythia{{\sc Pythia}}

\input epsf.tex
\def\desepsf(#1 width #2){\epsfxsize=#2 \epsfbox{#1}}
\def\kt{\ensuremath{k_t}}
\def\ktz{\ensuremath{k_{t,0}}}

\newcommand{\Pmax}{\mu^2}

\newcommand{\alphas}{\ensuremath{\alpha_\mathrm{s}}}

\newcommand{\PBM}{PB}
\newcommand{\TMDlib}{{\sc TMDlib}}
\newcommand{\TMDplotter}{{\sc TMDplotter}}
\newenvironment{tolerant}[1]{\par\tolerance=#1\relax}{ \par }
\usepackage{amsmath,bm}
\usepackage{lineno}

\usepackage{cite,./mcite}
\usepackage{tikz}
\usepackage[symbol]{footmisc}

\newcommand{\dglap}{Gribov:1972ri,Lipatov:1974qm,Altarelli:1977zs,Dokshitzer:1977sg}

\providecommand{\DOI}[1]{\href{http://dx.doi.org/#1}}

\begin{document}

\begin{flushright}
DESY 18-042 
\end{flushright}

\begin{center} {\sffamily\Large\bfseries
Collinear and TMD parton densities from fits to precision DIS measurements 
                           in the parton branching method }
 \\ \vspace{0.5cm}
{ \Large 
A.~Bermudez~Martinez$^{1}$,
P.~Connor$^{1}$,
F.~Hautmann$^{2,3,4}$,
H.~Jung$^{1}$,
A.~Lelek$^{1}$,
V.~Radescu$^{3,5}$\footnote[3]{Now at IBM Germany},
R.~\v{Z}leb\v{c}\'{i}k$^{1}$
}\\ \vspace*{0.15cm}
{\large $^1$DESY, Hamburg, FRG}\\
{\large $^2$RAL, Chilton OX11 0QX and University of Oxford, OX1 3NP} \\
{\large $^3$Elementary Particle Physics, University of Antwerp, B 2020  Antwerp}\\
{\large $^4$UPV/EHU, University of the Basque Country, E 48080 Bilbao}  \\
{\large $^5$CERN,  CH-1211 Geneva 23} \\
\end{center}
\begin{abstract}
Collinear and transverse momentum dependent (TMD) parton densities are obtained from  fits to precision measurements of   deep inelastic scattering (DIS)  cross sections at  HERA. The parton densities are evolved by DGLAP evolution with next-to-leading-order (NLO) splitting functions using 
the parton branching   method, allowing one to determine simultaneously collinear and TMD densities for all flavors over a wide range in
 $x$, $\mu^2$ and  $\kt$, relevant for predictions at the LHC.
The DIS cross section is computed from the parton densities using perturbative NLO coefficient functions.

Parton densities satisfying angular ordering conditions are presented. Two sets of parton densities are obtained, differing in the renormalization scale choice for the argument in the strong coupling  \alphas . This 
is taken to be either the evolution scale $\mu$ or the transverse momentum $q_t$. While both choices 
yield similarly good $\chi^2$ values for the fit to DIS measurements,  especially the gluon density turns 
out to differ between the two sets.  

The TMD densities are used to predict the transverse momentum spectrum of $Z$-- bosons at the LHC. 

\end{abstract}

\section{Introduction} 
\label{Intro}

Parton density functions (PDFs) play an essential role for  precise predictions  of  production 
processes in  hadronic collisions, obtained  from the  factorization of the cross sections 
 in hard-scattering process and 
PDFs, containing a non-perturbative input with perturbatively calculable evolution. The most advanced determination of parton densities come from the application of  DGLAP~\cite{\dglap} evolution with next-to-leading order (NLO)~\cite{Curci:1980uw,Furmanski:1980cm} and next-to-next-to-leading order (NNLO)~\cite{Vogt:2004mw,Moch:2004pa} 
splitting functions. The collinear parton densities as a function of the longitudinal momentum fraction $x$ and the evolution scale $\mu^2$ are obtained by several groups, for example ABM~\cite{Alekhin:2017kpj}, CTEQ~\cite{Dulat:2015mca}, HERAPDF~\cite{Abramowicz:2015mha}, NNPDF~\cite{Ball:2017nwa} and MSTW~\cite{Martin:2009iq,Martin:2012da}. The different  groups use the same DGLAP evolution, with ordering in virtuality and the same choice of the renormalization scale, but they differ in, for example,  the treatment of heavy flavors,  and the experimental data sets which are used for the determination of the starting distributions.

In Refs.~\cite{Hautmann:2017xtx,Hautmann:2017fcj} a new method, the Parton Branching method (PB), was introduced to treat DGLAP evolution. The method applies at exclusive level, and provides an iterative 
solution of  the evolution equations.    
It  agrees with the usual methods to solve the DGLAP equations for inclusive 
distributions, but it provides also additional features: in addition to the standard ordering in virtuality, angular ordering can be applied with the necessary change in the argument of $\alphas$~\cite{Amati:1980ch,Gieseke:2003rz}. The transverse momentum at every branching vertex can be calculated, leading to
a natural determination of the transverse momentum dependent (TMD) parton densities. 
The \PBM\ method uses the unitarity formulation of QCD evolution equations~\cite{Ellis:1991qj} 
and is  close in spirit to the works 
in~\cite{Jadach:2003bu,Placzek:2007xb,Tanaka:2003ck,Tanaka:2003gk,Hoche:2017iem,Hoche:2017hno}. 
As shown in Refs.~\cite{Hautmann:2017fcj,Zlebcik:2017suw}, 
 it  can be applied to NLO and NNLO splitting functions.

In this article we present a determination of collinear and TMD parton densities at NLO applying the \PBM\ method for the parton evolution. The initial parton distributions are determined from a fit to HERA I+II inclusive DIS cross section measurements~\cite{Abramowicz:2015mha}. An early fit was presented in Ref.~\cite{Hautmann:2017fcj}. Here, we present results obtained with angular ordering, both for collinear (integrated, iTMD) and TMD parton densities, and for different choices of the renormalization scale in \alphas\ including a full treatment of experimental and model dependent uncertainties.
We show an application of these TMDs to the  calculation of the transverse momentum of the $Z$-boson in Drell-Yan (DY) production at the Large Hadron Collider (LHC).

\section{Parton Branching method and evolution equation}
The \PBM\ method has been described in detail in Refs.~\cite{Hautmann:2017xtx,Hautmann:2017fcj}.  
Here we limit ourselves to recalling its  main elements. 

\subsection{General features}
The method is based on introducing a soft-gluon resolution scale  $z_M$ into the QCD evolution equations  to separate resolvable and non-resolvable emissions, and treating these via, respectively,   the resolvable splitting probabilities  $P_{ba}^{(R)} (\alphas , z ) $   and the Sudakov form factors  
\begin{equation}
\label{sud-def}
 \Delta_a ( z_M, \mu^2 , \mu^2_0 ) = 
\exp \left(  -  \sum_b  
\int^{\mu^2}_{\mu^2_0} 
{{d \mu^{\prime 2} } 
\over \mu^{\prime 2} } 
 \int_0^{z_M} dz \  z 
\ P_{ba}^{(R)}\left(\alphas , 
 z \right) 
\right) 
  \;\; .   
\end{equation}
Here 
$a , b$ are flavor indices, 
$\alphas$ is the strong coupling at a scale being a function of ${\mu}^{\prime 2}$ to be specified in Section \ref{pdf_fits},   
$z$ is 
the longitudinal momentum 
splitting variable, and   
$z_M < 1 $ is the soft-gluon resolution parameter.  For easier reading we use the notation  $ \Delta_a ( \mu^2 ) = \Delta_a ( z_M, \mu^2 , \mu^2_0 ) $.
The form factors  eq.~(\ref{sud-def}) have the interpretation of probabilities  for non-resolvable branchings between the evolution scales $\mu_0$ and $\mu$. The functions  $P_{ba}^{(R)} (\alphas , z ) $ have the structure 
\begin{equation}
P_{ba}^{(R)} (\alphas ,z) = 
 \delta_{ba}  k_b (\alphas) \ 
{ 1 \over {  1 - z  }} 
+ R_{ba} (\alphas ,z)
\; ,   
\label{realPba}
\end{equation}
where the first term on the right hand side contains the pole singularity in the soft-gluon  radiation region $z \to 1$ and the second term contains  logarithmic terms and analytic terms for  $z \to 1$. The coefficients $k_b$ and $ R_{ba}$ in eq.~(\ref{realPba}) have the perturbation  series expansions 
\begin{equation}
k_{b} (\alphas) = \sum^{\infty}_{n=1} \left( \frac{\alphas}{2\pi}
\right)^{n} k_{b}^{(n-1)}  
\, , \;\; 
R_{ba} (\alphas ,z) = \sum^{\infty}_{n=1} \left( \frac{\alphas}{2\pi}
\right)^{n} R_{ba}^{(n-1)}(z)
\, . 
\label{Aab}
\end{equation}

The explicit expressions 
for the $ n = 1 $ (LO) 
and $n=2$ (NLO)  contributions in the expansions in
eq.~(\ref{Aab}) are given  
in~\cite{Hautmann:2017fcj}.  
The  $n = 3$ (NNLO) contributions can be read   from~\cite{Vogt:2004mw,Moch:2004pa}  and are used for NNLO calculations in the \PBM\ method in~\cite{Zlebcik:2017suw}. 
 The integrals appearing in the Sudakov form factors eq.~(\ref{sud-def}) 
are positive at LO, NLO and NNLO, while the 
 functions eq.~(\ref{realPba})  
 can be negative  at NLO and NNLO.
The positivity of the integrals in eq.~(\ref{sud-def}) is essential for the application of the \PBM\ method.

The \PBM\ method  allows one 
to   take   
into account simultaneously 
soft-gluon emission in the region 
$z \to 1$  and transverse momentum 
$ {\bf q}_\perp$ recoils in the parton branchings along the QCD cascade. 
Its advantage is twofold: on one 
hand, in collinear distributions 
additional QCD features can be 
studied such as the color radiation's 
angular ordering, determined by  
soft-gluon interferences, and 
its effects on factorization and renormalization scales; on the other hand, the 
method  can be applied to obtain 
transverse momentum dependent 
(TMD) 
distributions. 

  The \PBM\ evolution  equations for 
 TMD parton  densities    
$ {\cal A}_a ( x , {\bf k } , \mu^2) $ 
are given 
by~\cite{Hautmann:2017fcj}   
\begin{eqnarray}
\label{evoleqforA}
   { {\cal A}}_a(x,{\bf k}, \mu^2) 
 &=&  
 \Delta_a (  \mu^2  ) \ 
 { {\cal A}}_a(x,{\bf k},\mu^2_0)  
 + \sum_b 
\int
{{d^2 {\bf q}^{\prime } } 
\over {\pi {\bf q}^{\prime 2} } }
 \ 
{
{\Delta_a (  \mu^2  )} 
 \over 
{\Delta_a (  {\bf q}^{\prime 2}  
 ) }
}
\ \Theta(\mu^2-{\bf q}^{\prime 2}) \  
\Theta({\bf q}^{\prime 2} - \mu^2_0)
 \nonumber\\ 
&\times&  
\int_x^{z_M} {{dz}\over z} \;
P_{ab}^{(R)} (\alphas 
,z) 
\;{ {\cal A}}_b\left({x \over z}, {\bf k}+(1-z) {\bf q}^\prime , 
{\bf q}^{\prime 2}\right)  
  \;\;  ,     
\end{eqnarray}
in terms of the   $\Delta_a$ form factors, eq.~(\ref{sud-def}),  and   $P_{ba}^{(R)}$  functions, eq.~(\ref{realPba}). The scale in \alphas\ is a function of ${\bf q}^{\prime 2}$, as discussed in Section~\ref{pdf_fits}. These equations can be  solved by an iterative Monte Carlo method. In this method every resolvable branching is reconstructed explicitly and the full kinematics at each branching is taken into account.  The \PBM\ method allows to solve eq.~(\ref{evoleqforA}) in an easy and direct way, with the possibility to include, for example, also heavy quark masses and soft-gluon coherence conditions.
 
The collinear parton densities $ { f}_a(x,\mu^2) $ are related to the TMD densities by   
\begin{equation} 
\label{unintA}
 { f}_a(x,\mu^2) = 
\int  
 \ {\cal A}_a ( x , {\bf k } , \mu^2)  
\  { {d^2 {\bf k }} \over \pi} 
  \; ,  
\end{equation}
and are  described  as integrated TMD (iTMD). The  evolution equations  for iTMD densities analogous to eq.~(\ref{evoleqforA}) can be written as 
\begin{equation}
\label{sudintegral2}
   f_a(x,\mu^2)  =   \Delta_a ( \mu^2 ) \ f_a(x,\mu^2_0)  
+  \sum_b \int^{\mu^2}_{\mu^2_0} {{d \mu^{\prime 2} } \over \mu^{\prime 2} } {
{\Delta_a ( \mu^2 )} 
 \over 
{\Delta_a ( \mu^{\prime 2}
 ) }
}\int_x^{z_M} {{dz}\over z} P_{ab}^{(R)} \left(z,\alphas \right) 
\; f_b\left({\frac{x}{z}},\mu^{\prime 2}\right)      \;\; .   
\end{equation}
These  equations  have been  shown  to be 
equivalent to DGLAP evolution equations at NLO~\cite{Jadach:2003bu,Placzek:2007xb,Hautmann:2017fcj,Hautmann:2017xtx} and NNLO~\cite{Zlebcik:2017suw} for $\alpha_s=\alpha_s(\mu^{\prime 2})$ and $z_M \to 1$.

\subsection{\PBM\ method and determination of initial distribution}
The \PBM\ method has been implemented in the \verb+xFitter+ package \cite{Alekhin:2014irh} to allow  fits to be made  to  cross section measurements. A full Monte Carlo solution of the evolution equation for every new set of initial parameters would be too time consuming to be efficient. Instead, a method developed already in \cite{Hautmann:2013tba,Jung:2012hy,Hautmann:2014uua}   is applied:
first,  a kernel $ {\cal K}^{int}_{b a}\left(x'',\mu_0^2,\Pmax\right) $  is determined from the Monte Carlo  solution of the evolution equation for any initial parton\footnote{In practice, since the initial state partons can be only light quarks or gluons, it is enough to determine the kernel $ {\cal  K}$  only for one  initial  state quark and a gluon.}
 of flavor $b$ evolving to a final parton of flavor $a$;  then 
this kernel  is folded with the non-perturbative starting 
distribution $f_{0,b} (x,\mu_0^2)$,   
\begin{eqnarray}
xf_a(x,\mu^2) 
 &= &x\int dx' \int dx'' f_{0,b} (x',\mu_0^2) {\cal  K}^{int}_{ba}\left(x'',\mu_0^2,\Pmax\right) 
 \delta(x' 
x'' - x) 
\nonumber  
\\
& = & \int dx' {f_{0,b} (x',\mu_0^2) }  
\frac{x}{x'} \ { {\cal  K}^{int}_{ba}\left(\frac{x}{x'},\mu_0^2,\Pmax\right) }  \;\; .
\label{iTMD_kernel}
\end{eqnarray}
The kernel  ${\cal  K}^{int}_{ba}$ includes the full parton evolution from $\mu_0^2$ to $\mu^2$, as in eq.~(\ref{sudintegral2}), with  Sudakov form factors and splitting probabilities,  and  is determined with the \PBM\ method.  In eq.~(\ref{iTMD_kernel}) 
the kernel ${\cal K}^{int}_{ba}$ depends on $x$, $\mu_0^2$ and $\Pmax$ for the $\kt$-integrated (iTMD) distributions. 

To include also the transverse momentum $\kt$,  we define a new kernel $ {\cal K}_{b a}\left(x'',\ktz^2,\kt^2,\mu_0^2,\Pmax\right) $  for the TMD   distributions, with $k_t^2 = {{\bf k }^2  }$,

\begin{eqnarray}
x{\cal A}_a(x,\kt^2,\mu^2) 
 &= &x\int dx' \int dx'' {\cal A}_{0,b} (x',\ktz^2,\mu_0^2) {\cal  K}_{ba}\left(x'',\ktz^2,\kt^2,\mu_0^2,\Pmax\right) 
 \delta(x' x'' - x) 
\nonumber  
\\
& = & \int dx' {\cal A}_{0,b} (x',\ktz^2,\mu_0^2)
\frac{x}{x'} \ { {\cal  K}_{ba}\left(\frac{x}{x'},\ktz^2,\kt^2,\mu_0^2,\Pmax\right) }  \;\; .
\label{TMD_kernel}
\end{eqnarray}
The evolution  of the kernel starts at $x_0=1$ at $\mu_0^2$. 
In general, the starting distribution ${\cal A}_0$ can have flavor and $x$ dependent $\ktz$ distributions, for simplicity we use here 
a factorized form:
\begin{eqnarray}
{\cal A}_{0,b} (x,\ktz^2,\mu_0^2) & = & f_{0,b} (x,\mu_0^2) \cdot \exp(-| \ktz^2 | / \sigma^2)
\label{TMD_A0}
\end{eqnarray}
where the intrinsic $\ktz$ distribution is given by a Gauss distribution  with $ \sigma^2  =  q_0^2 / 2 $  for all flavors and all $x$ with a constant value $q_0 = 0.5$~\GeV. 

Technically, the results of the kernel evolution are stored in a grid of size $50 \times 50 ( \times 50)$ (for the TMD densities). The grid spacing is logarithmic ($\mu_0 < \mu < 14000~\GeV$ and $0.01 <  \kt < 14000~\GeV$), the $x$ range is divided into 5 subregions with logarithmic spacing: subregions of 10 bins are defined with the boundaries $10^{-6}, 0.01, 0.1, 0.4, 0.9, 1$ which is optimized to ensure appropriate behavior for large $x$, where the parton densities (and the kernel) are varying rapidly.

\begin{figure}[htb]
\begin{center} 
\includegraphics[width=0.47\textwidth]{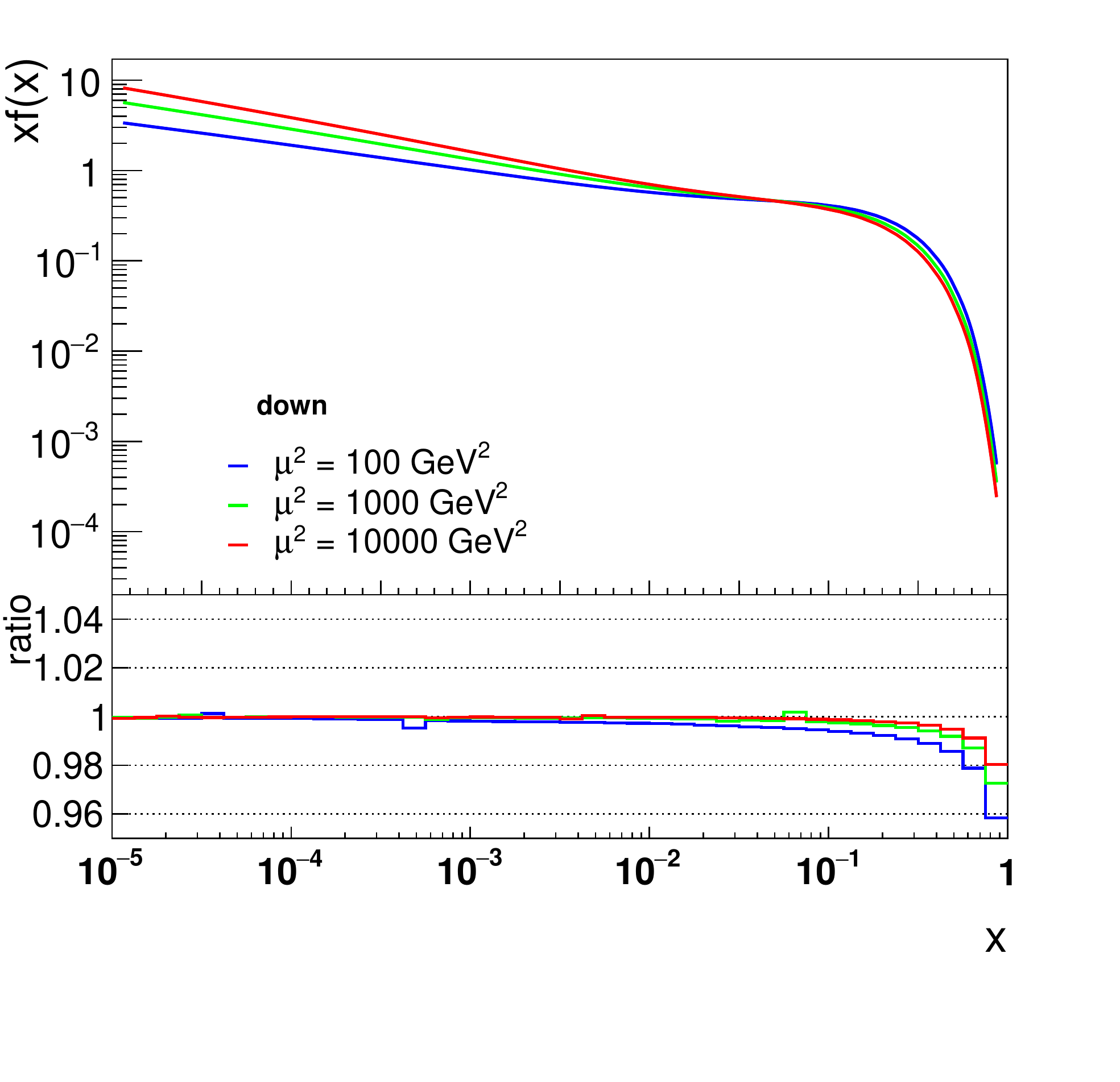} \hskip 0.5cm
\includegraphics[width=0.47\textwidth]{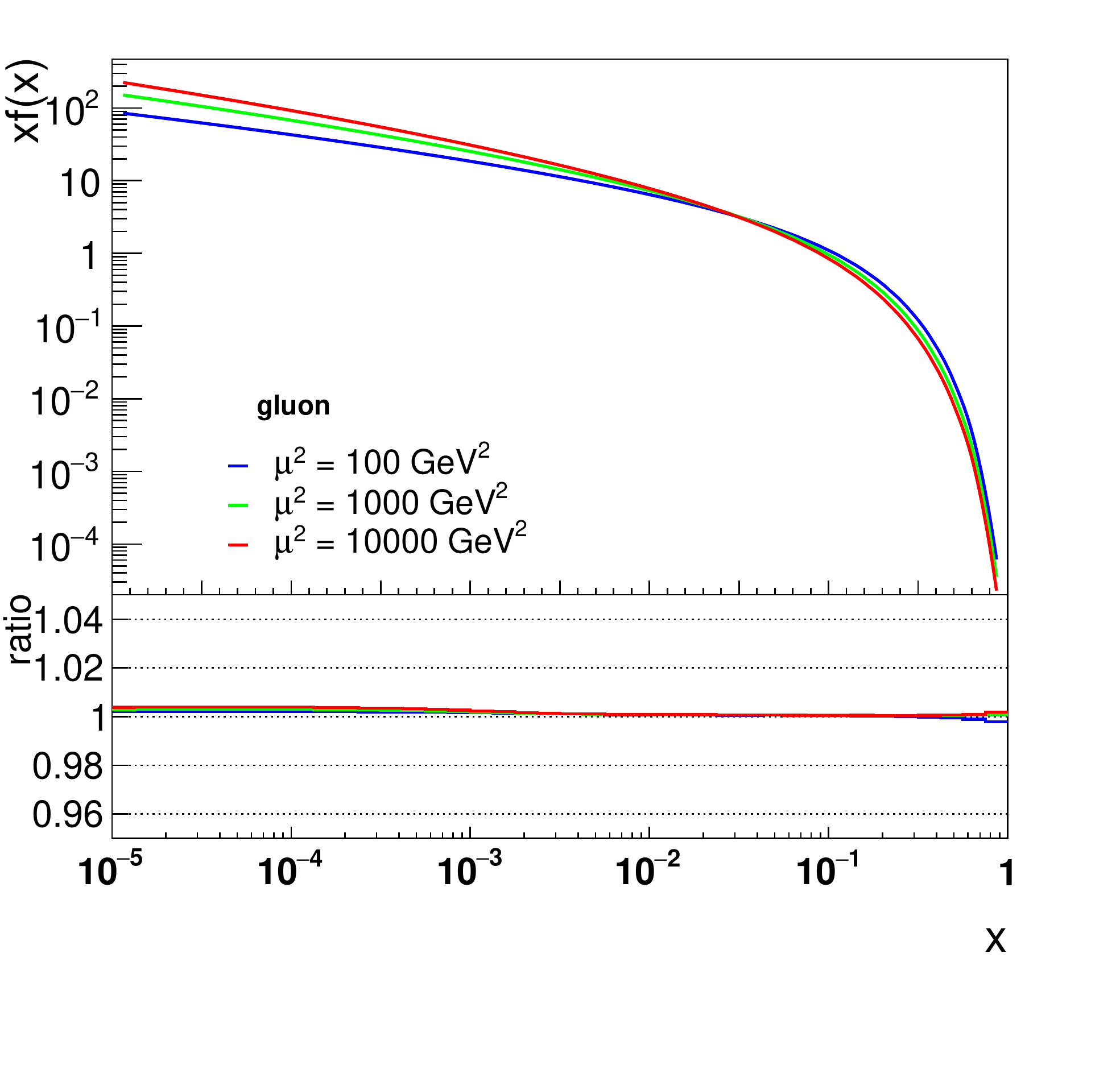} 
  \caption{\small Comparison of the results from the convolution in eq.(\protect\ref{iTMD_kernel}) with the prediction from QCDNUM~\protect\cite{Botje:2010ay} using the same input distributions, for d-quarks (left) and gluons (right) at different values of the evolution scale $\mu^2$ starting from $\mu_0^2=1.9~\protect\GeV^2$ with $\alphas(\mu^{\prime 2})$. The lower panels show the ratio of the parton density with the one predicted by QCDNUM. The evolution is performed with NLO DGLAP splitting functions and using $z_M = 0.99999$. }
\label{convolution}
\end{center}
\end{figure}

In Fig.~\ref{convolution} we show the result of  convoluting the starting distribution (here taken to be 
the benchmark parameterization of Ref.~\cite{Giele:2002hx}) with the kernel as given in eq.~(\ref{iTMD_kernel}) for the integrated distribution,  and compare this  with the prediction from a standard evolution program (QCDNUM) for different values of the evolution scale $\mu^2$. The kernel is evolved using NLO splitting functions with  resolution scale parameter $z_M$, separating resolvable from non-resolvable branchings, set to the 
value   $z_M = 0.99999$.  
Very good agreement is observed over the whole range. Only 
the quark distribution shows differences at very large $x$ of the order of a few percent, which come from the finite grid spacing in $x$ when storing the kernel (changing to a uniform logarithmic grid spacing in $x$ leads to significantly larger deviations at large $x$). 
In most of the phase space region relevant for high precision physics at HERA and the LHC the differences are at the  per mille level.

\section{Parton densities obtained from fits to inclusive HERA DIS measurements}
\label{pdf_fits}
The most recent and most precise measurements of the  lepton-proton DIS cross section over a wide range in $x$ and $Q^2$ were performed at HERA with a combination of the measurements from the H1 and ZEUS collaborations~\cite{Abramowicz:2015mha}. These measurements are the basis for any determination of parton densities. In Ref.~\cite{Abramowicz:2015mha} a fit to the inclusive DIS measurements was performed using DGLAP at LO, NLO and NNLO, resulting in the HERAPDF2.0 parton distributions. These fits were performed with QCDNUM~\cite{Botje:2010ay} within the \verb+xFitter+ framework~\cite{Alekhin:2014irh}  using a starting scale $\mu_0 = 1.9$~GeV$^2$ and the renormalization and factorization scales set to $\mu^2_r = \mu^2_f = Q^2$. The light quark matrix elements were taken from QCDNUM, the heavy-quark contributions were obtained within  the general-mass variable-flavor scheme RTOPT~\cite{Thorne:2012az,Thorne:2006qt,Thorne:1997ga} for neutral current, while for charged current interactions the zero-mass approximation from QCDNUM was used. 
The mass of the charm quark is set  $m_c=1.47$~GeV, and $m_b=4.5$~GeV is used for the bottom quark mass. The strong coupling is set to $\alpha_s (M_z^2)  =0.118$. 

The parameterized PDFs are the gluon distribution, $xg$,
the valence-quark distributions, $xu_v$, $xd_v$, and
the $u$-type and $d$-type anti-quark distributions,
$x\bar{U}$, $x\bar{D}$. The relations $x\bar{U} = x\bar{u}$ and
$x\bar{D} = x\bar{d} +x\bar{s}$ are assumed at the starting scale $\mu_{0}$.

The following parameterizations are used for the different parton flavors:
 \begin{eqnarray}
xg(x) &=   & A_g x^{B_g} (1-x)^{C_g}  - A'_g x^{B'_g} (1-x)^{C'_g},  \nonumber \\
xu_v(x) &=  & A_{u_v} x^{B_{u_v}}  (1-x)^{C_{u_v}}\left(1+E_{u_v}x^2 \right) ,  \nonumber \\
xd_v(x) &=  & A_{d_v} x^{B_{d_v}}  (1-x)^{C_{d_v}} ,  \nonumber \\
x\bar{U}(x) &=  & A_{\bar{U}} x^{B_{\bar{U}}} (1-x)^{C_{\bar{U}}}\left(1+D_{\bar{U}}x\right) ,  \nonumber \\
x\bar{D}(x) &= & A_{\bar{D}} x^{B_{\bar{D}}} (1-x)^{C_{\bar{D}}} .
\label{HERAPDF_paramtrization}
\end{eqnarray}

The quark-number sum rules and the momentum sum rule can be used to constrain the normalization parameters, $A_{u_v}, A_{d_v}, A_g, A'_g$. 
The $B$ parameters  are set $B_{\bar{U}}=B_{\bar{D}}$ for the sea distributions. The strange-quark distribution is parameterized as a $d$-type sea with an $x$-independent fraction, $f_s$, $x\bar{s}= f_s x\bar{D}$ at $\mu^2_{\rm_{0}}$ with $f_s=0.4$. A further constraint was applied by setting $A_{\bar{U}}=A_{\bar{D}} (1-f_s)$. 
 \begin{tolerant}{2000}
A total of 1145 data points of neutral-current and charged-current deep-inelastic cross section measurements were used in the range of
$ 3.5 < Q^2 < 50000 $ GeV$^2$ and \mbox{$4\cdot 10^{-5}< x < 0.65 $}. 
\end{tolerant}
The same data sets, kinematic ranges and  hard-scattering coefficient functions,  including the heavy-quark treatment,  are used for the fits described here. 
We use NLO DGLAP splitting functions~\cite{Curci:1980uw,Furmanski:1980cm} as well as  NLO coefficient functions~\cite{Furmanski:1981cw} for light quarks. For heavy quarks we apply the general-mass variable-flavor scheme RTOPT~\cite{Thorne:2012az,Thorne:2006qt,Thorne:1997ga} for neutral current, while for charged current interactions the zero-mass approximation is used. 

In the next section we determine the free parameters of the  initial distributions given by 
eq.~(\ref{HERAPDF_paramtrization})  via  
 fits to the HERA DIS data in the range of  $Q^2>3.5$~GeV$^2$ using NLO DGLAP splitting functions within the \PBM\ method using $z_M=0.99999$. 

\begin{figure}[htb]
\begin{center} 
\includegraphics[width=0.17\textwidth]{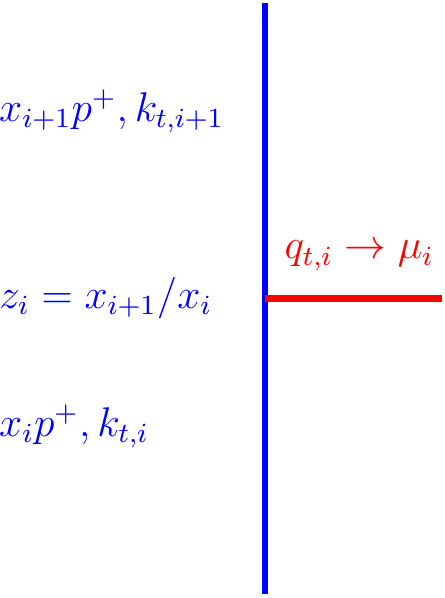} \hskip 2cm
\includegraphics[width=0.25\textwidth]{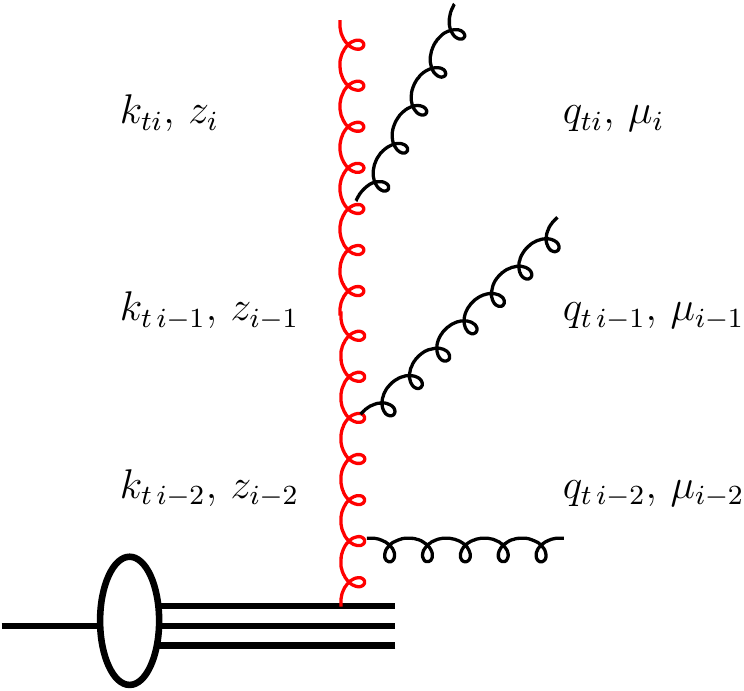} 
  \caption{\small Left: Branching process $ b \to a + c$. Right: Schematic view of a parton branching process.}
\label{parton-branching}
\end{center}
\end{figure} 

\begin{table}[htb]
\renewcommand*{\arraystretch}{1.0}
\centerline{
\begin{tabular}{|c|c|c|c|}
\hline
\multicolumn{4}{|c|}{PB NLO Set1 $\alphas(\mu_i^2)$ } \\
\hline
\multicolumn{1}{|c|}{ } &
\multicolumn{1}{c|}{$\chi^2$} &
\multicolumn{1}{c|}{d.o.f} &
\multicolumn{1}{c|}{$\chi^2/$d.o.f} \\
\hline
$\mu^2_{0}= 1.9$~GeV$^2$   & 1363.37 & 1131  &   1.21   \\
\hline
\hline
\multicolumn{4}{|c|}{PB NLO Set 2 $\alphas(q_{t\,i}^2)$ } \\
\hline
\multicolumn{1}{|c|}{ } &
\multicolumn{1}{c|}{$\chi^2$} &
\multicolumn{1}{c|}{d.o.f} &
\multicolumn{1}{c|}{$\chi^2/$d.o.f} \\
\hline
$\mu^2_{0}= 1.4$~GeV$^2$   & 1369.80   & 1131  &   1.21  \\
\hline
\end{tabular}}
\caption{\small Values of $\chi^2$ for the different fits at NLO.} 
\label{Fit_chi2}
\end{table}

The \PBM\ method allows the explicit calculation of the kinematics at every branching vertex (see Fig.~\ref{parton-branching} left). 
Once the physical meaning of the evolution scale is specified in terms of kinematic variables,
the transverse momenta of the propagating and  emitted partons  can be calculated. 
In Ref.~\cite{Hautmann:2017xtx} it was pointed out that   angular ordering  gives 
 transverse momentum distributions which are stable with respect to variations of the resolution parameter  $z_M$.
 In angular ordering, the angles of the emitted partons increase from the hadron side towards the hard scattering, as shown in Fig.~\ref{parton-branching} right.
The transverse momentum $q_{t\,i}$ can be calculated  in terms of the angle $\Theta_i$ of the emitted parton with respect to 
 the beam directions from $q_{t,i} = (1-z_i) E_{i} \sin \Theta_i$. Associating the "angle"  $E_i \sin \Theta_i$ with $\mu_i$ gives 
\begin{equation}
  \label{ang-ordering}
 {\bf q}_{t,i}^2  =  (1-z_i)^2 \mu_i^ 2  \;\; .
\end{equation}

\begin{figure}[htb]
\begin{center} 
\includegraphics[width=0.3\textwidth]{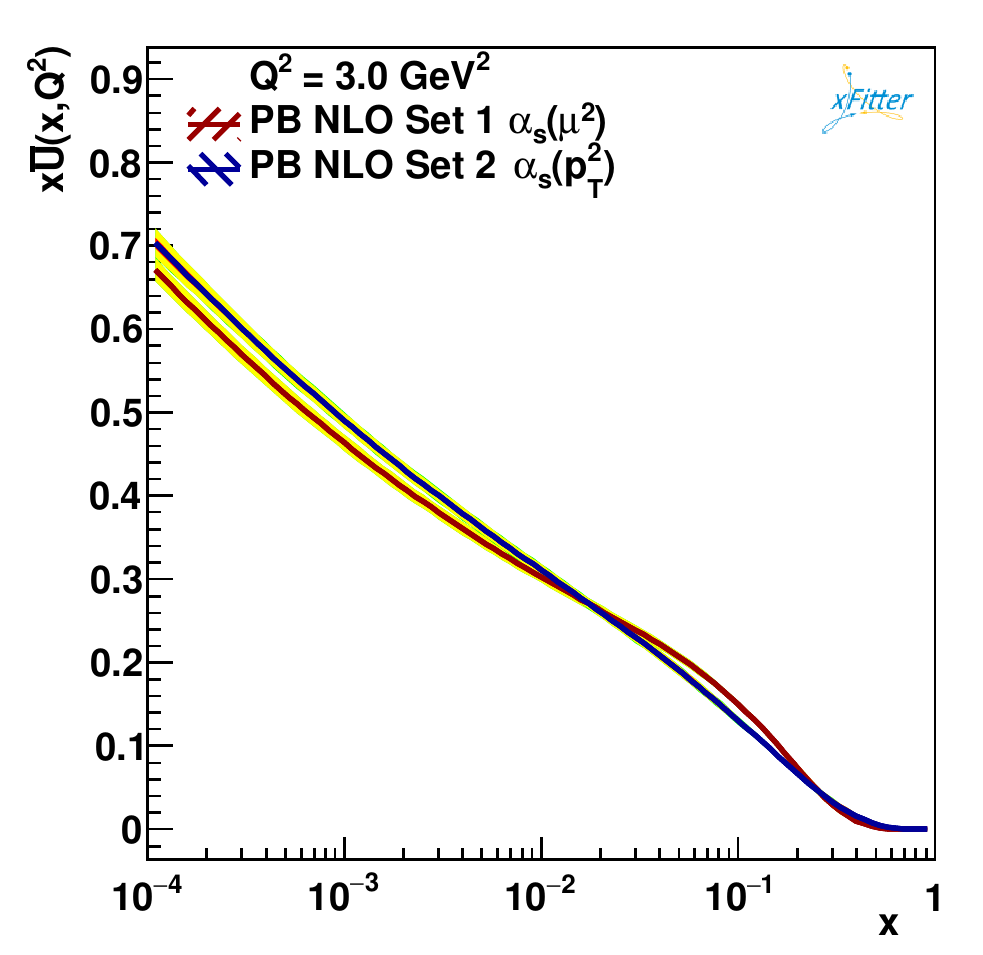} 
\includegraphics[width=0.3\textwidth]{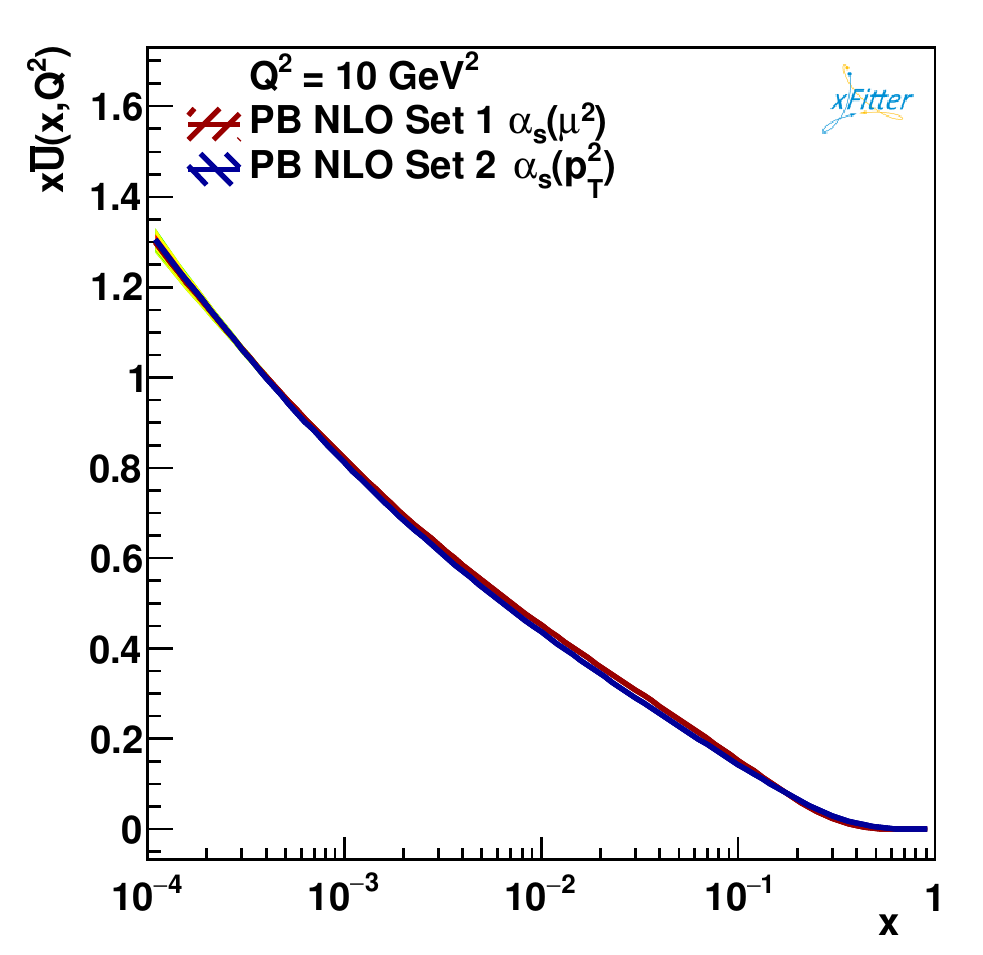} 
\includegraphics[width=0.3\textwidth]{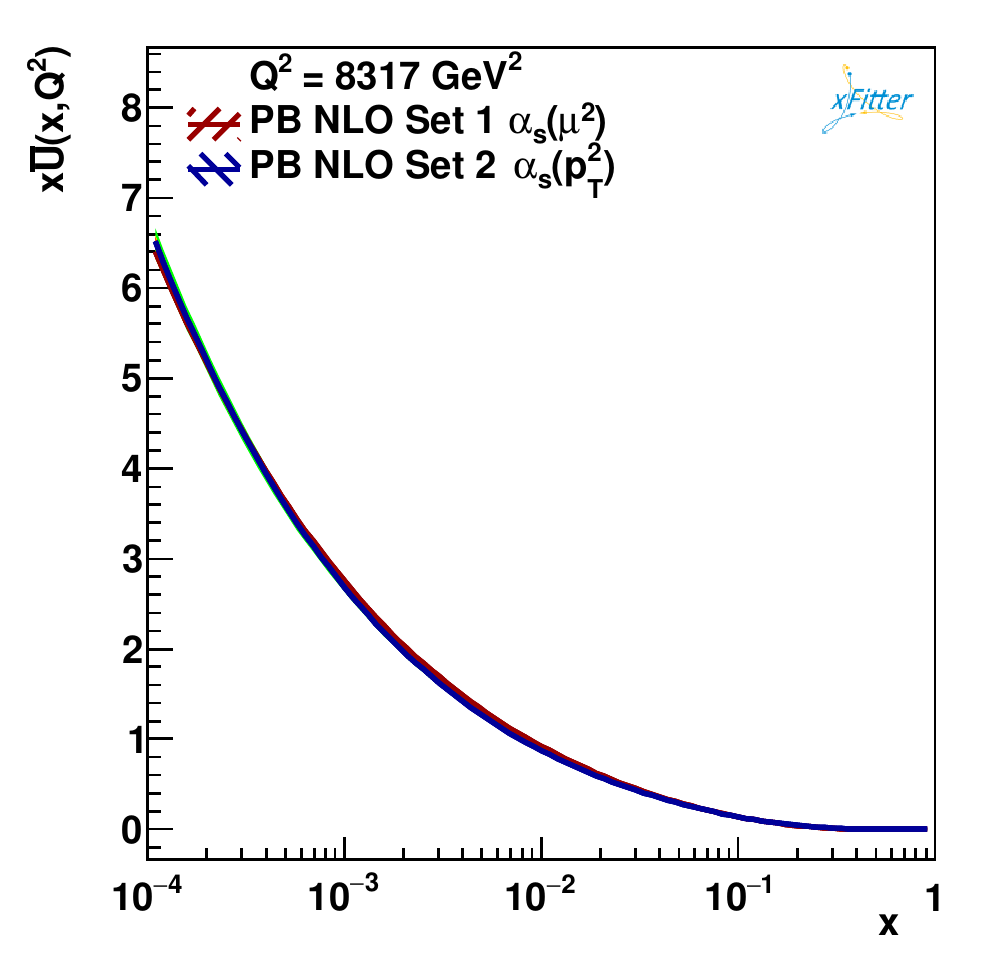} 
\includegraphics[width=0.3\textwidth]{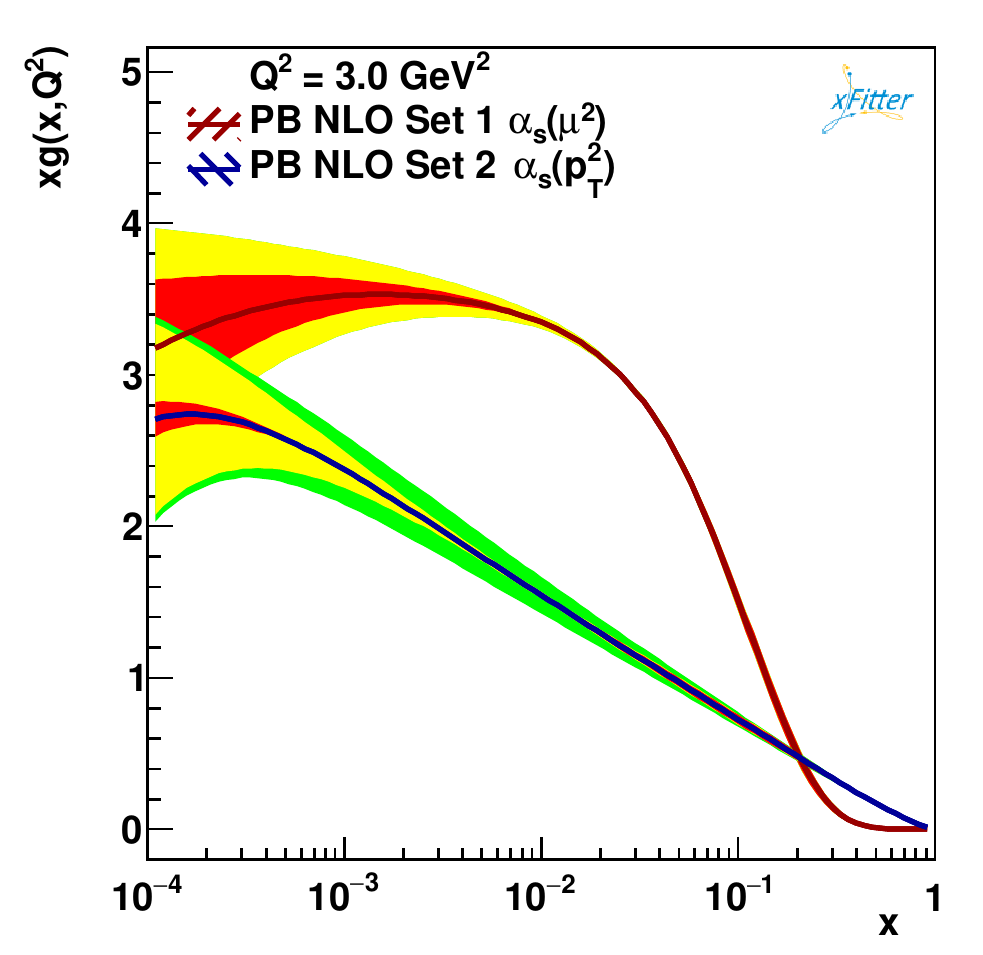} 
\includegraphics[width=0.3\textwidth]{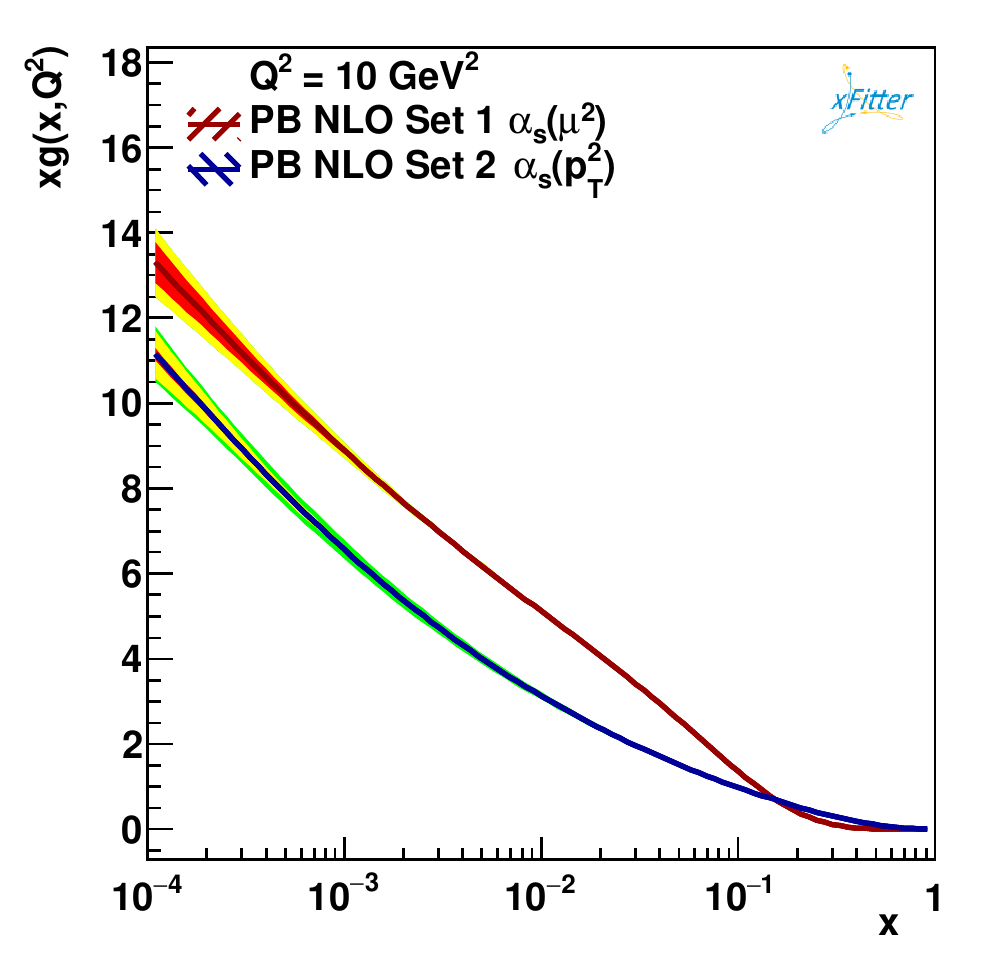} 
\includegraphics[width=0.3\textwidth]{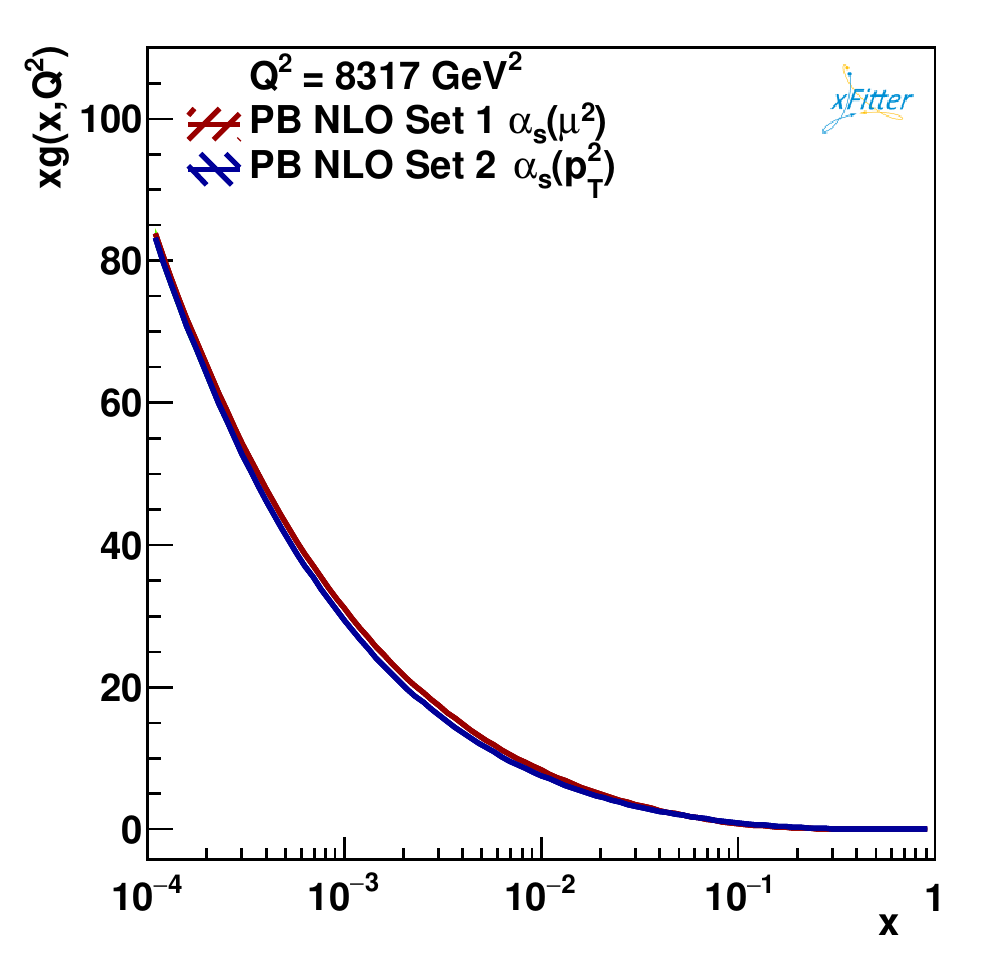} 
  \caption{\small Parton densities for different values of the scale $\mu^2=Q^2$.
  The different choices for the renormalization scale in \alphas\ are shown. 
  The red band shows the experimental uncertainty, the yellow band the model dependence. The green band shows the uncertainty coming from the variation of the parameter $q_{cut}$ in Set~2.}
\label{collinear_pdfs}
\end{center}
\end{figure} 

In the following, we use the \PBM\ method to determine  collinear (iTMD)  and transverse momentum  dependent (TMD) parton densities using  NLO DGLAP splitting functions for two different scenarios: first we only apply the angular ordering condition  for the  calculation of the transverse momentum and  keep the evolution scale $\mu^2_i$ as the argument in $\alphas$ (Set~1); in a second scenario (Set~2), we use (in eqs.~(\ref{sud-def},\ref{evoleqforA},\ref{sudintegral2})) the transverse momentum $|{\bf q}^2_{t,i}|$ as the argument in \alphas , as suggested in Ref.~\cite{Amati:1980ch,Gieseke:2003rz}. An additional parameter $q_{cut}$ needs to be introduced in $\alphas(\max(q^2_{cut},|{\bf q}_{t,i}^2|))$ to avoid the non-perturbative region, since with large $z$ the scale  $|{\bf q}_{t,i}^2|  =  (1-z_i)^2 \mu_i^ 2$ can become very small.  We take the default choice for this parameter to be 
 $q_{cut}=1$~GeV,  and we estimate the model dependence with a variation around the default choice.

In the first case, the integrated parton density, and the initial parameters, will be the same (up to numerical precision) as the ones obtained by HERAPDF2.0, and we use this as a benchmark for the whole method. In the second case, even the integrated parton distributions differ, because of the different scale in \alphas. In both cases a reasonably good fit is obtained with $\chi^2/ndf \sim 1.2$, as for HERAPDF2.0. In Tab.~\ref{Fit_chi2} results of the fits are given. The starting scale $\mu_0^2$ is chosen differently for the 2 scenarios:
for Set~1 we chose (as in HERAPDF) $\mu_0^2 = 1.9$~GeV$^2$ while for Set~2 we chose $\mu_0^2 = 1.4$~GeV$^2$, which gave the best 
 $\chi^2/ndf $. In the appendix we show results obtained from a fit when $\mu_0^2 = 1.9$~GeV$^2$ is chosen instead of $\mu_0^2 = 1.4$~GeV$^2$. The distributions agree within their uncertainties.
The values of the parameters at the starting scale $\mu_0^2$ are given in Tab.~\ref{tab:paramNLO}.

\begin{table}[htbp]
\renewcommand*{\arraystretch}{1.0}
\centerline{
\begin{tabular}{|l|l|r|l|l|l|l|l|l|l|}
\hline
\multicolumn{9}{|c|}{PB NLO Set1 $\alphas(\mu_i^2)$ } \\
\hline 
\multicolumn{1}{|c|}{ } &
\multicolumn{1}{c|}{~~~$A$~~~} &
\multicolumn{1}{c|}{~~~$B$~~~} &
\multicolumn{1}{c|}{~~~$C$~~~} &
\multicolumn{1}{c|}{~~~$D$~~~} &
\multicolumn{1}{c|}{~~~$E$~~~} &
\multicolumn{1}{c|}{~~~$A'$~~~} &
\multicolumn{1}{c|}{~~~$B'$~~~} &
\multicolumn{1}{c|}{~~~$C'$~~~} \\
\hline
$xg$      & 4.32   & $-$0.015  & 9.15 &  &   & 1.040 & $-$0.166 & 25 \\
$xu_v$     & 4.07   &  0.714  & 4.84 &  & 13.5 & & & \\
$xd_v$     & 3.15   &  0.806  & 4.07 &  &  & &  & \\
$x\bar{U}$ & 0.107 & $-$0.173 & 8.05 & 11.8 &   & & & \\
$x\bar{D}$ & 0.178 & $-$0.173 & 4.89 &   &  & & & \\
\hline \hline
\multicolumn{9}{|c|}{PB NLO Set 2 $\alphas(q_{t\,i}^2)$ } \\
\hline
$xg$      & 0.42   & $-$0.047  & 0.96 &  &   & 0.008 & $-$0.58 & 25 \\
$xu_v$     & 2.49   &  0.65  & 3.44 &  & 13.7 & & & \\
$xd_v$     & 2.02  &  0.75  & 2.47 &  &  & &  & \\
$x\bar{U}$ & 0.14 & $-$0.16 & 5.29 & 1.5 &   & & & \\
$x\bar{D}$ & 0.24 & $-$0.16 & 5.83 &   &  & & & \\
\hline 
\end{tabular}}
\caption{\small Parameter values of the initial distributions at NLO. The parameter $C'=25$ was fixed, as in HERAPDF2.0.
The parameters correspond to a starting scale $\mu_0^2 = 1.9 (1.4)$ GeV$^2$  for Set 1 (Set 2). }
 \label{tab:paramNLO}\end{table}

\subsection{Collinear Parton Densities (iTMD)} 
\label{sec:iTMD}

The fits to HERA measurements are performed using   $\chi^2$ minimization, as in the case of  the HERAPDF fits, implemented in \verb+xFitter+  \cite{Alekhin:2014irh}. The definition of $\chi^2$ includes systematic shifts, a treatment of correlated and uncorrelated systematic uncertainties. In total 162 systematic uncertainties plus procedural uncertainties from the combination of H1 and ZEUS are treated as correlated uncertainties. 
 \begin{figure}[h!tb]
\begin{center} 
\includegraphics[width=0.3\textwidth]{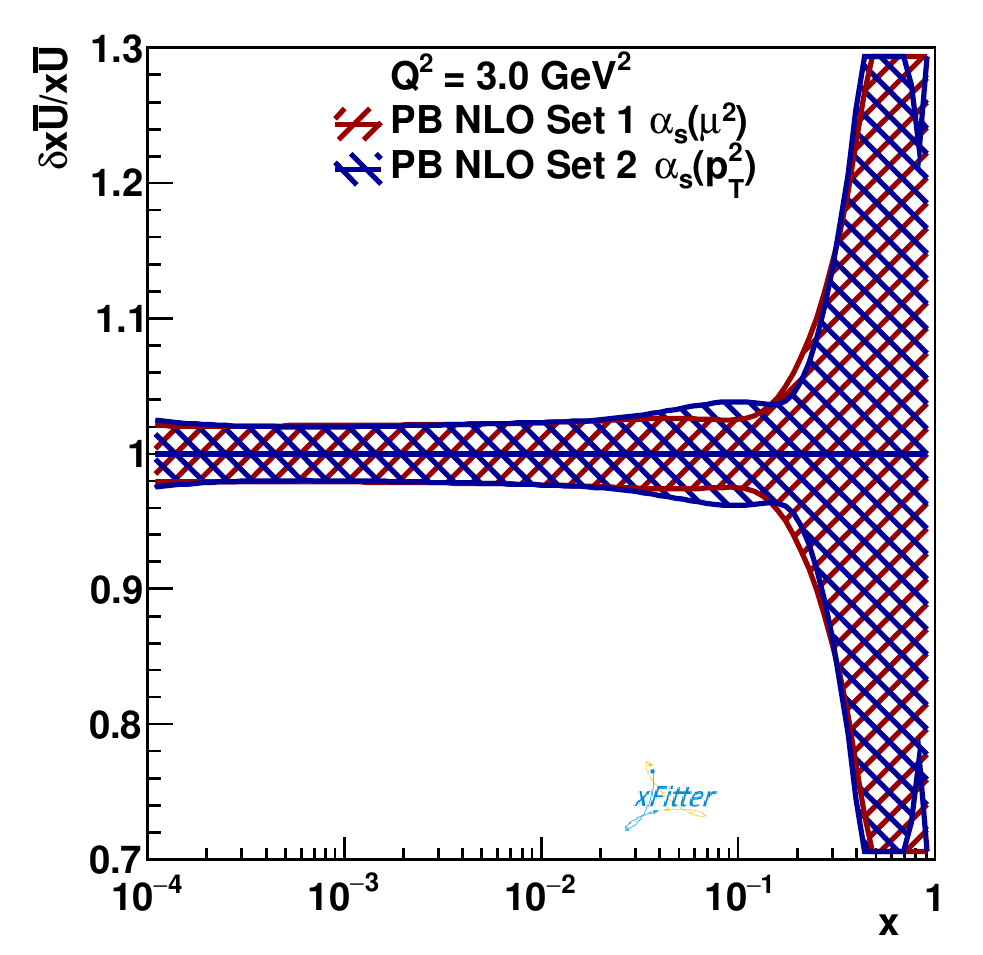}  
\includegraphics[width=0.3\textwidth]{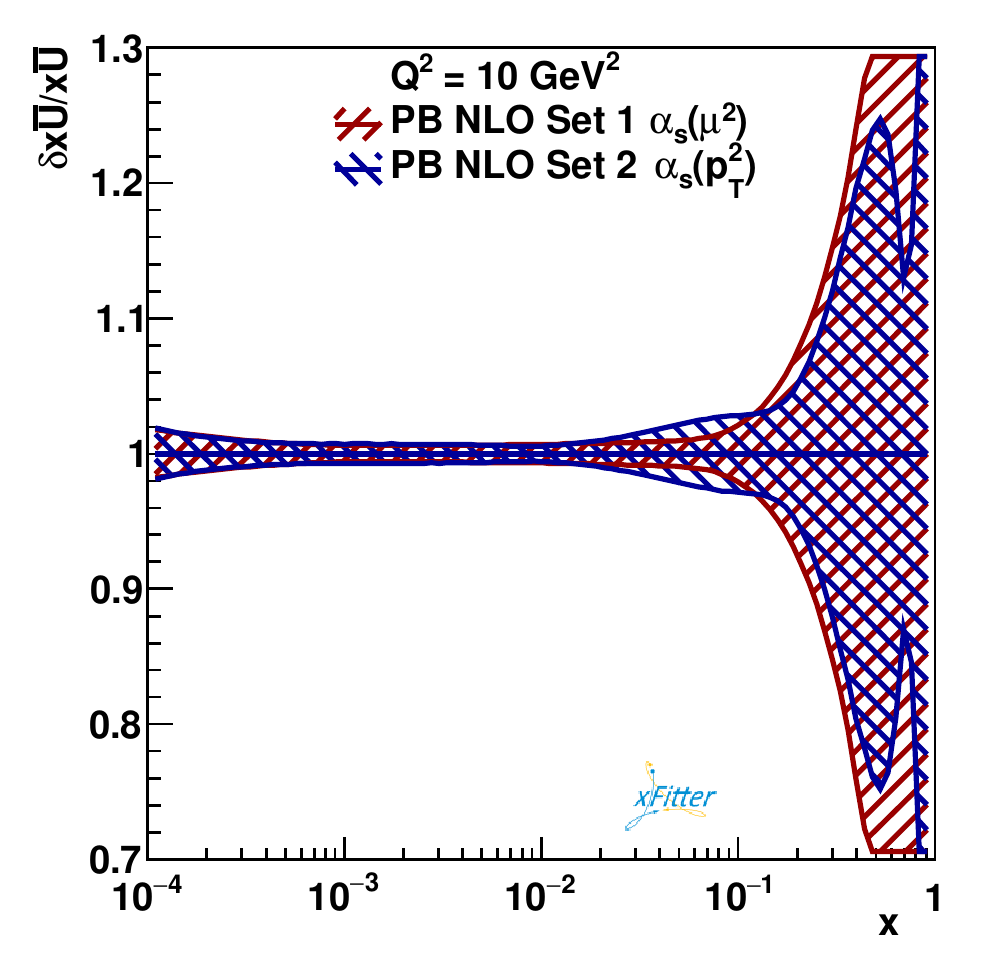}  
\includegraphics[width=0.3\textwidth]{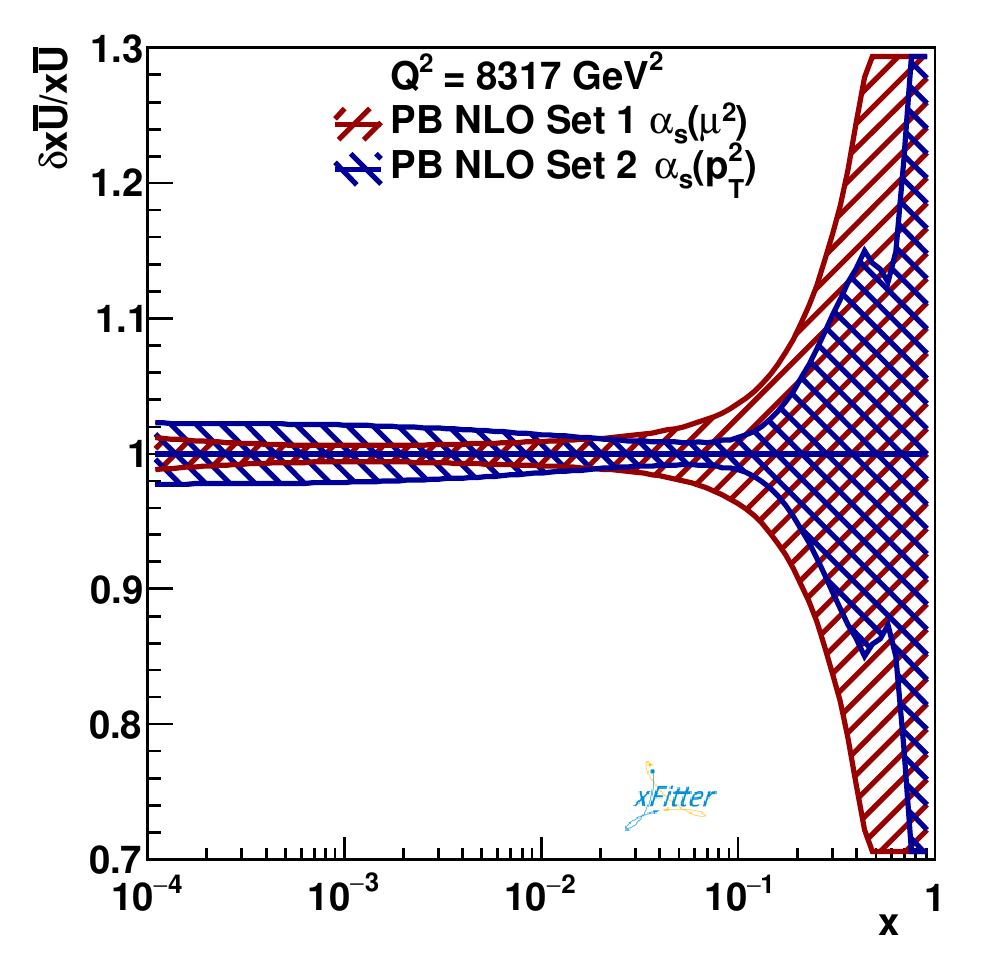}  
\includegraphics[width=0.3\textwidth]{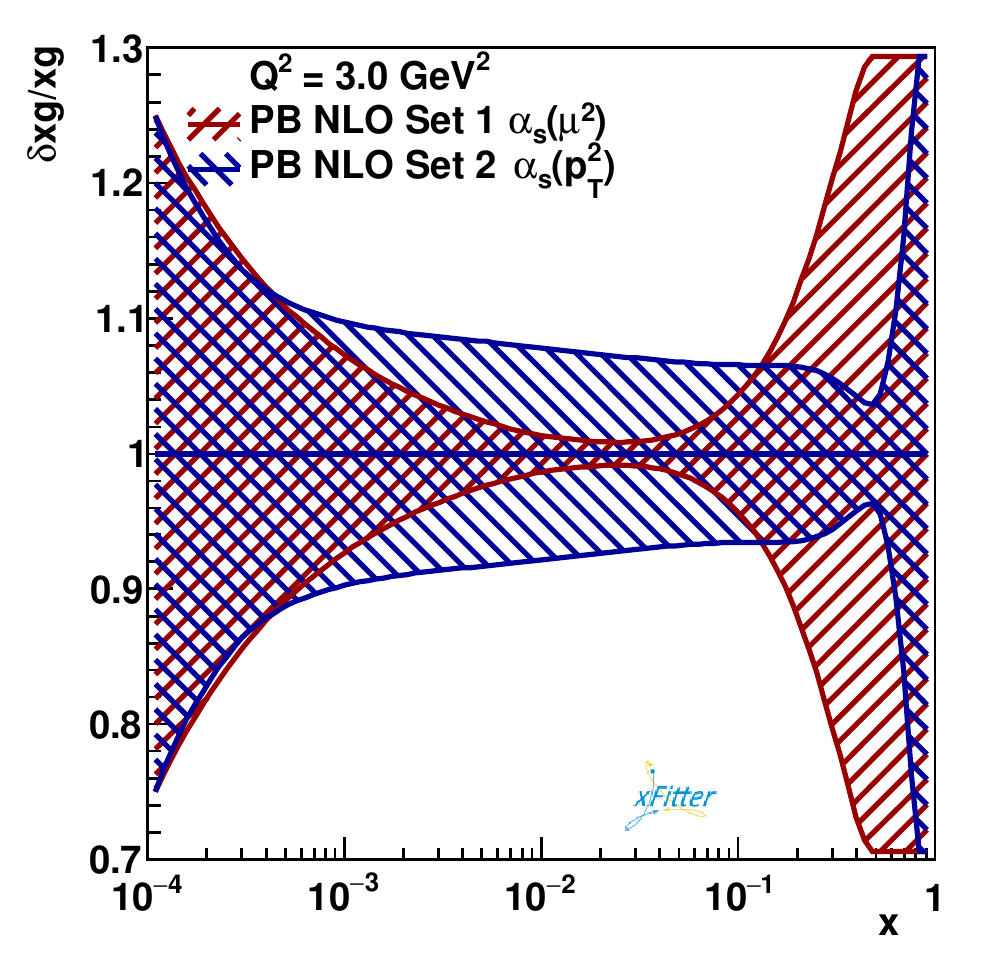}  
\includegraphics[width=0.3\textwidth]{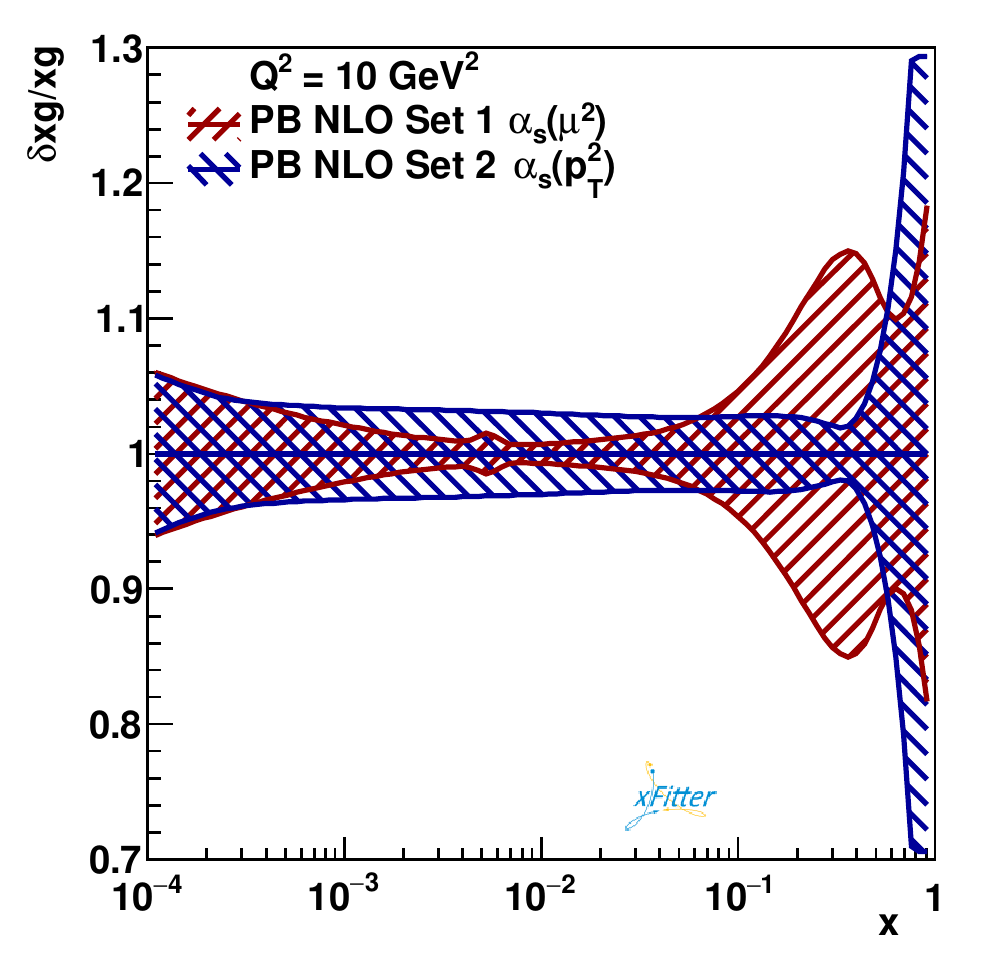}  
\includegraphics[width=0.3\textwidth]{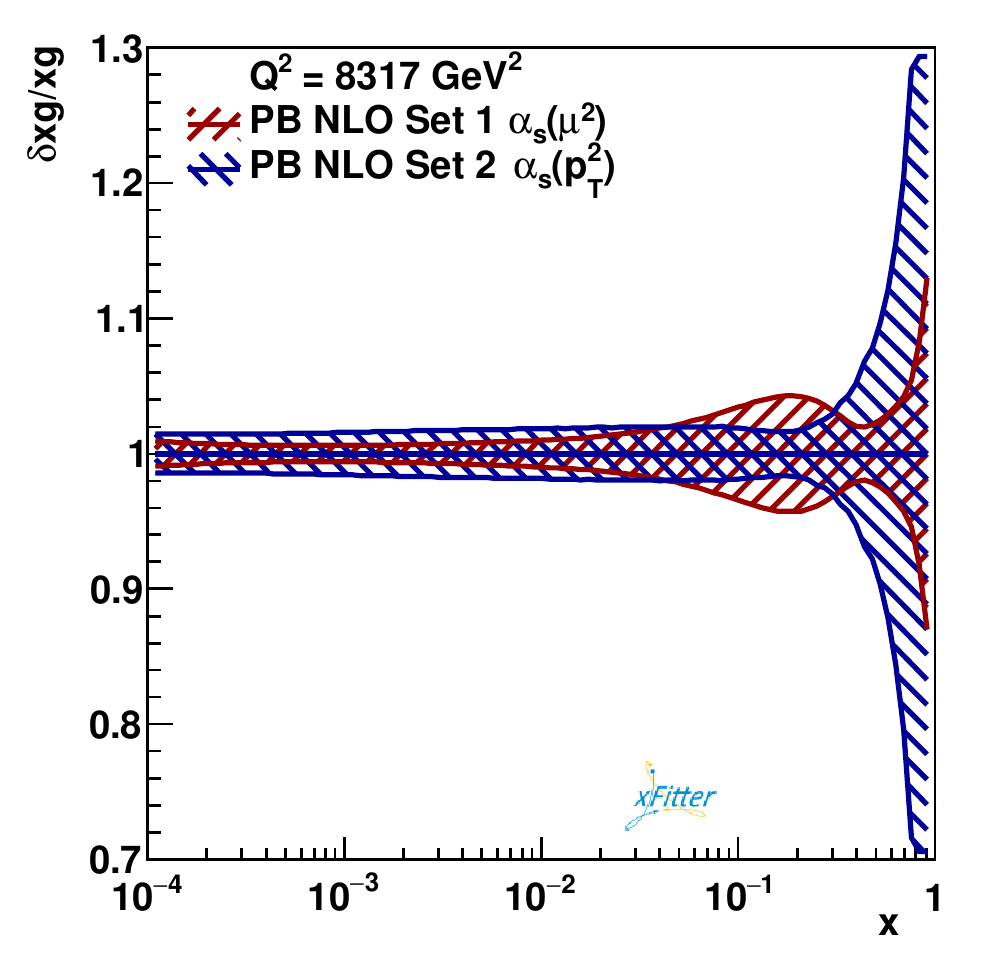}  
  \caption{\small Total uncertainties (experimental and model uncertainties) for the two different sets at different values of the evolution scale $\mu^2$.}
\label{collinear_pdfs_ratio}
\end{center}
\end{figure} 
\begin{table}[htb]
\renewcommand*{\arraystretch}{1.0}
\centerline{
\begin{tabular}{|c|c|c|c|}
\hline
\multicolumn{1}{|l|}{ } &
\multicolumn{1}{c|}{Central } &
\multicolumn{1}{c|}{Lower} &
\multicolumn{1}{c|}{Upper}  \\
\multicolumn{1}{|c|}{ } &
\multicolumn{1}{c|}{value} &
\multicolumn{1}{c|}{value} &
\multicolumn{1}{c|}{value}  \\
\hline
PB NLO Set1  $\mu^2_0$ (GeV$^2$)	      & 1.9  & 1.6  & 2.2 \\
PB NLO Set 2  $\mu^2_0$ (GeV$^2$)	      & 1.4  & 1.1  & 1.7 \\
PB NLO Set 2  $q_{cut}$ (GeV)	      & 1.0  & 0.9  & 1.1 \\
$m_c$  (GeV)          & 1.47 & 1.41 & 1.53 \\
$m_b$  (GeV)         & 4.5 & 4.25 & 4.75 \\
\hline
\end{tabular}}
\caption{\small Central values and change ranges of parameters for model dependence} 
\label{Fit_model}
\end{table}
 
The experimental uncertainties of the resulting parton densities are determined with the Hessian method \cite{Pumplin:2001ct} (as implemented in \verb+xFitter+ ) with $\Delta \chi^2 = 1$.
The model dependence of the PDF fits is obtained by varying 
charm and bottom masses and the starting scale of the evolution $\mu_0^2$. For Set~2 also the parameter $q_{cut}$ is varied.
The central values and the range of variation is given in Tab.~\ref{Fit_model}.

In Fig.~\ref{collinear_pdfs} the $\bar U$-type quark and gluon densities are shown as  functions 
 of $x$ for different values of the evolution scale $\mu^2=Q^2$ including the experimental uncertainties (red band) and the uncertainties coming from the model dependence (yellow band). For Set~2 the uncertainty of the parameter $q_{cut}$ is shown as the green band. 
The results of Set~1 are identical to the ones obtained in HERAPDF 2.0.  Although the fits (Set~1 and Set~2) to HERA I+II data are of similar quality, the resulting parton distributions, especially for the gluon, are significantly different. 
With increasing evolution scale, however, they become more and more similar.

In Fig.~\ref{collinear_pdfs_ratio} the total uncertainties (experimental and model) of the parton densities are shown. The uncertainties of Set~2 for the gluon distribution at large $x$ become large.
We have investigated a possible bias coming from the chosen form of the parameterization by including additional terms for the gluon density:\begin{equation}
xg(x) =    A_g x^{B_g} (1-x)^{C_g} (1 + D_g x + E_g x^2)  - A'_g x^{B'_g} (1-x)^{C'_g}. \nonumber
\end{equation}
The obtained $\chi^2$ of the fit does change by at most 1 unit,  the resulting gluon distribution does not change visibly. Details of the bias study are given in the appendix.
\begin{figure}[h!tb]
\begin{center} 
\includegraphics[width=0.3\textwidth]{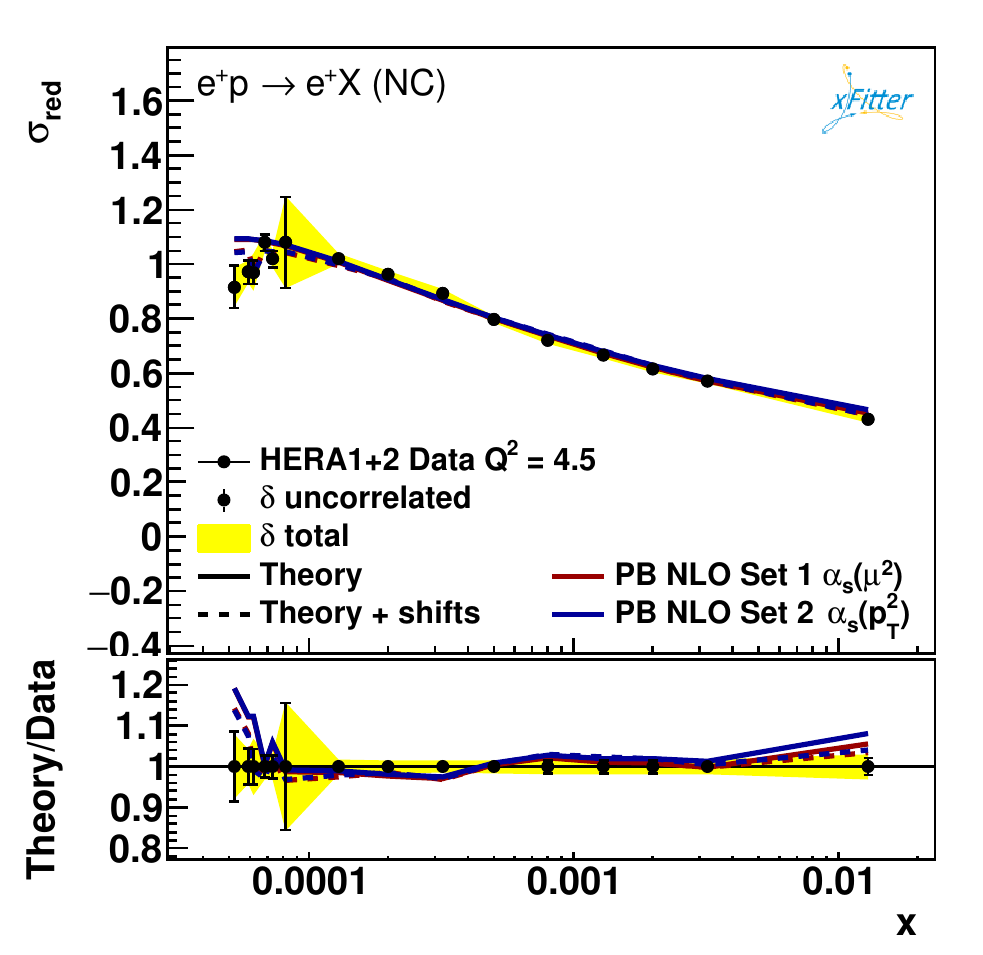}  
\includegraphics[width=0.3\textwidth]{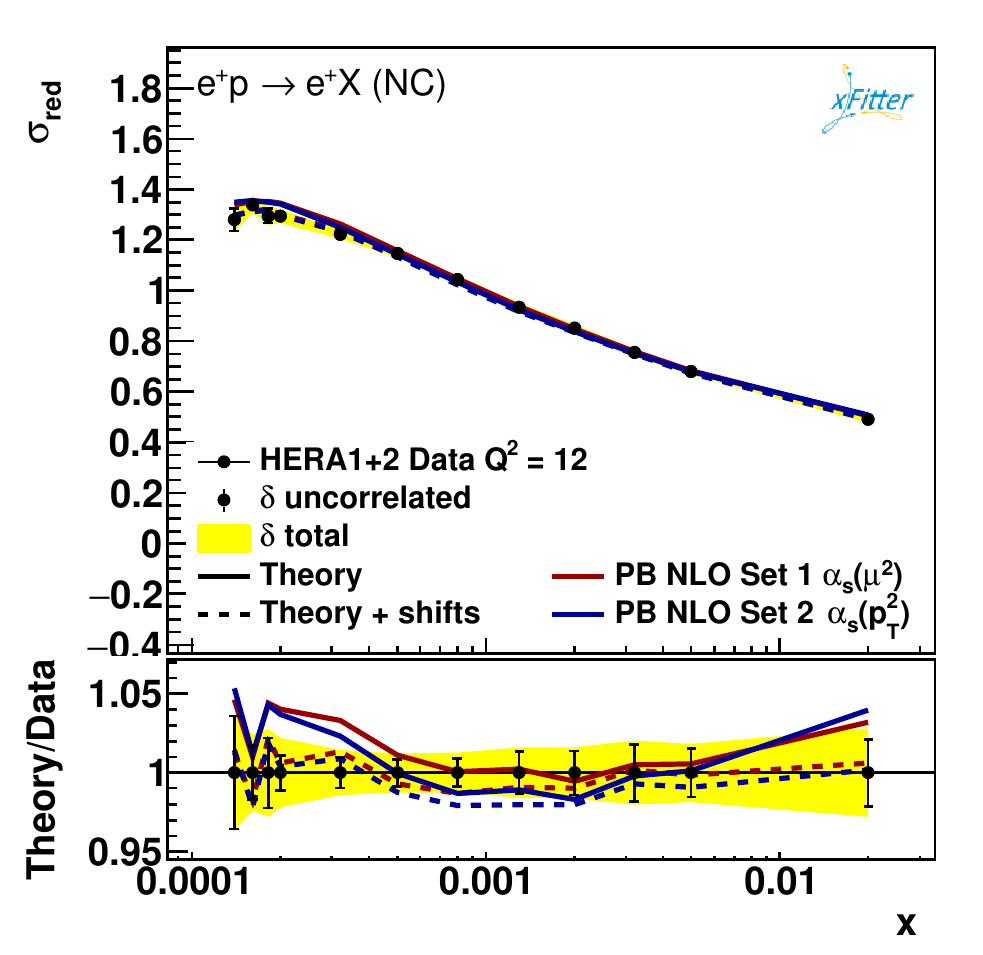}  
\includegraphics[width=0.3\textwidth]{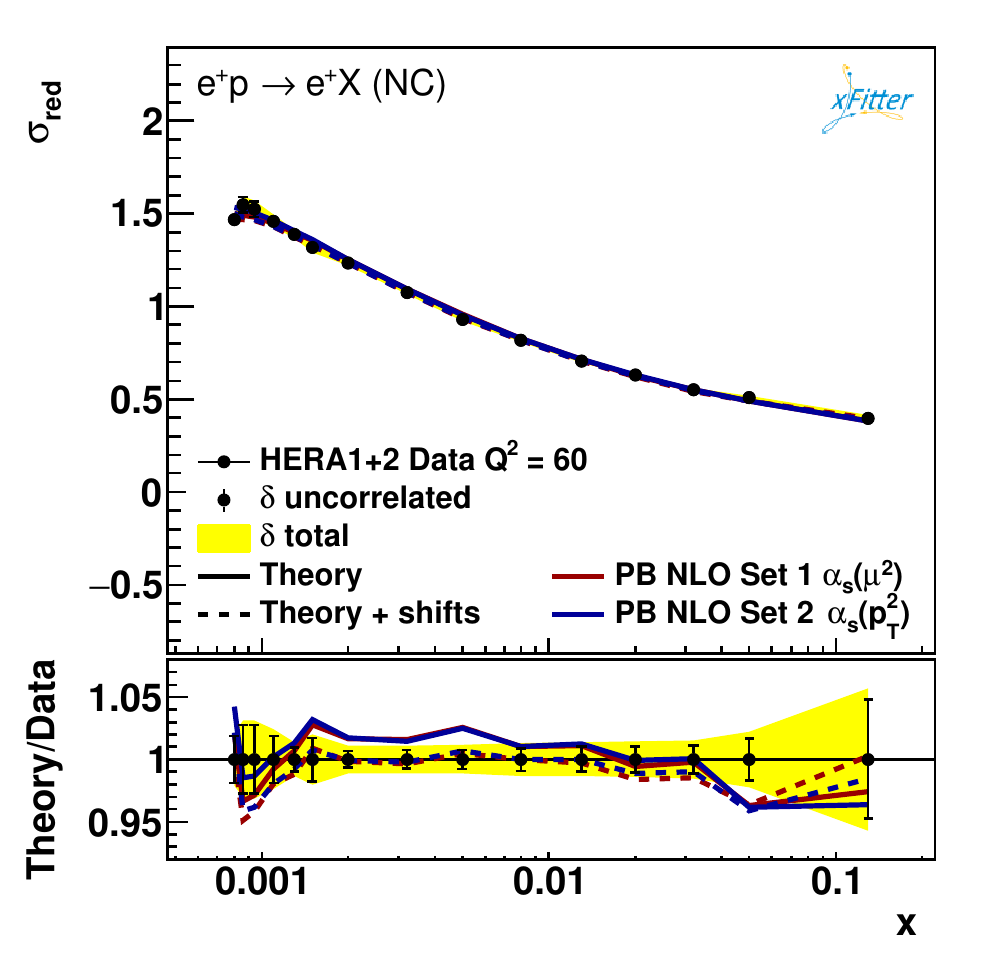}  
\includegraphics[width=0.3\textwidth]{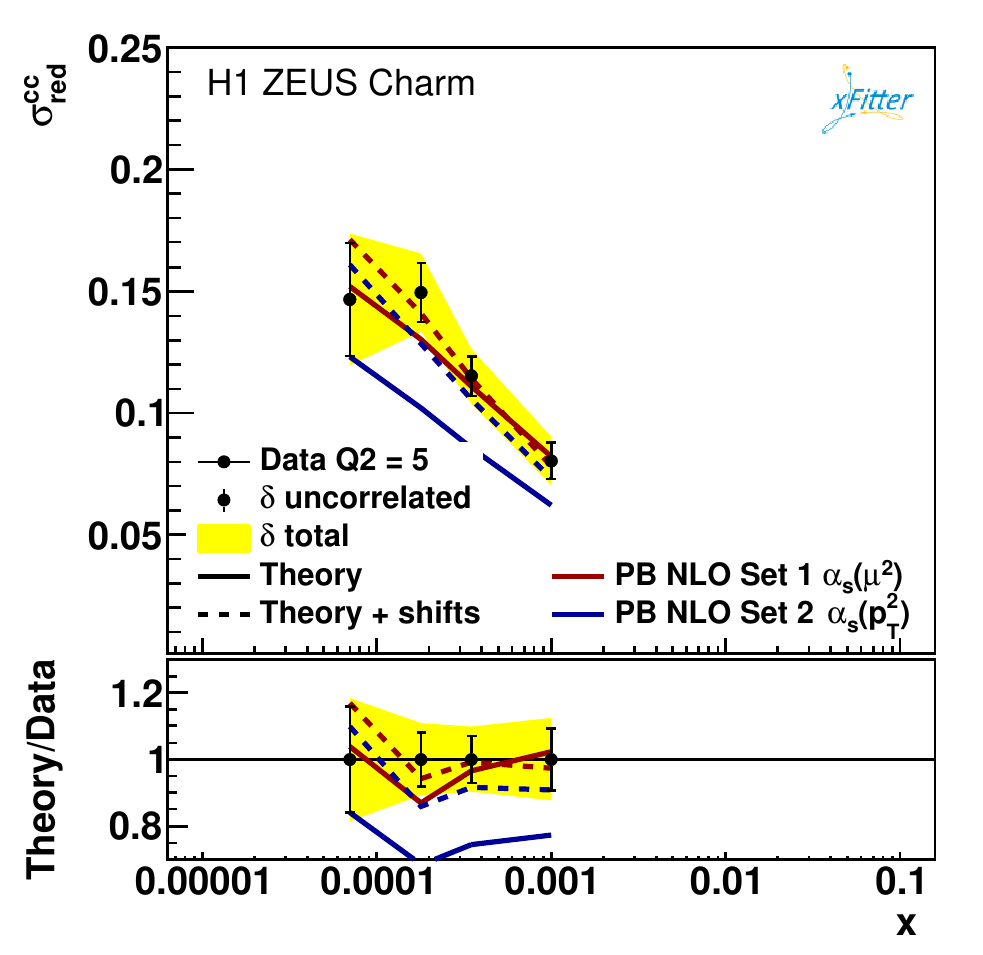}  
\includegraphics[width=0.3\textwidth]{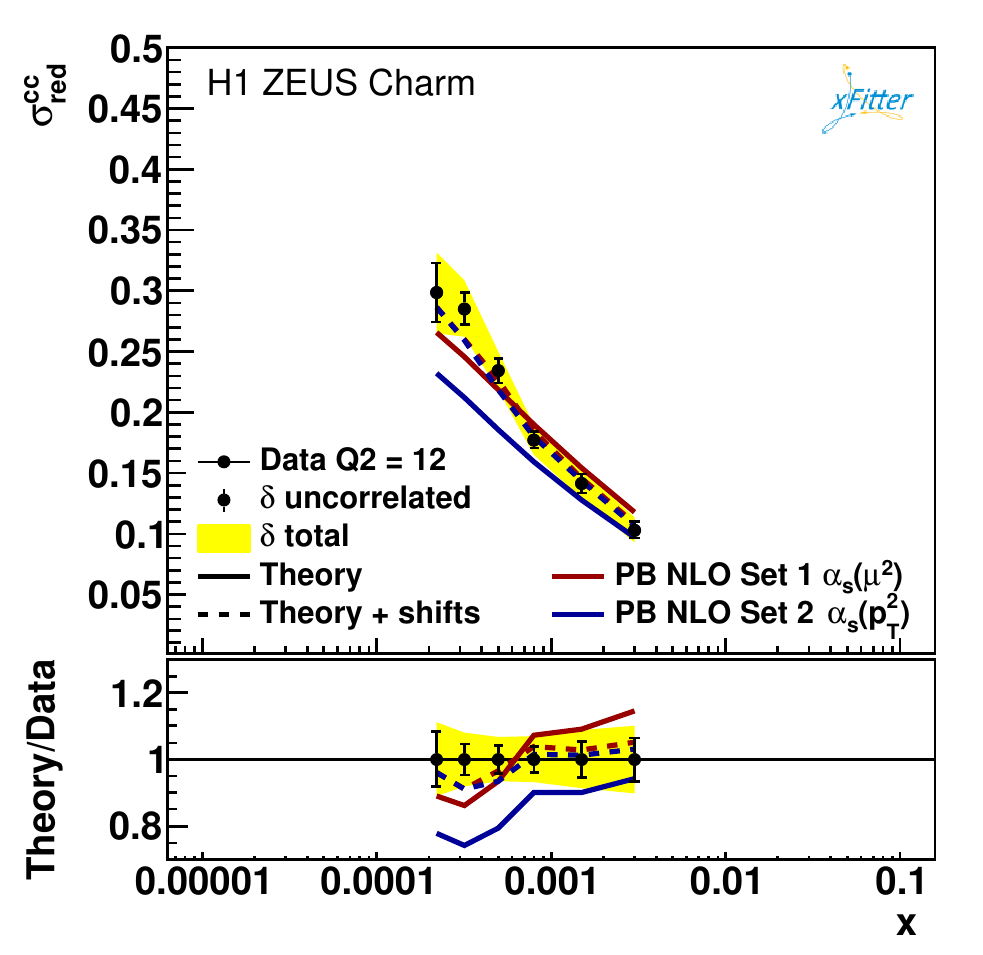}  
\includegraphics[width=0.3\textwidth]{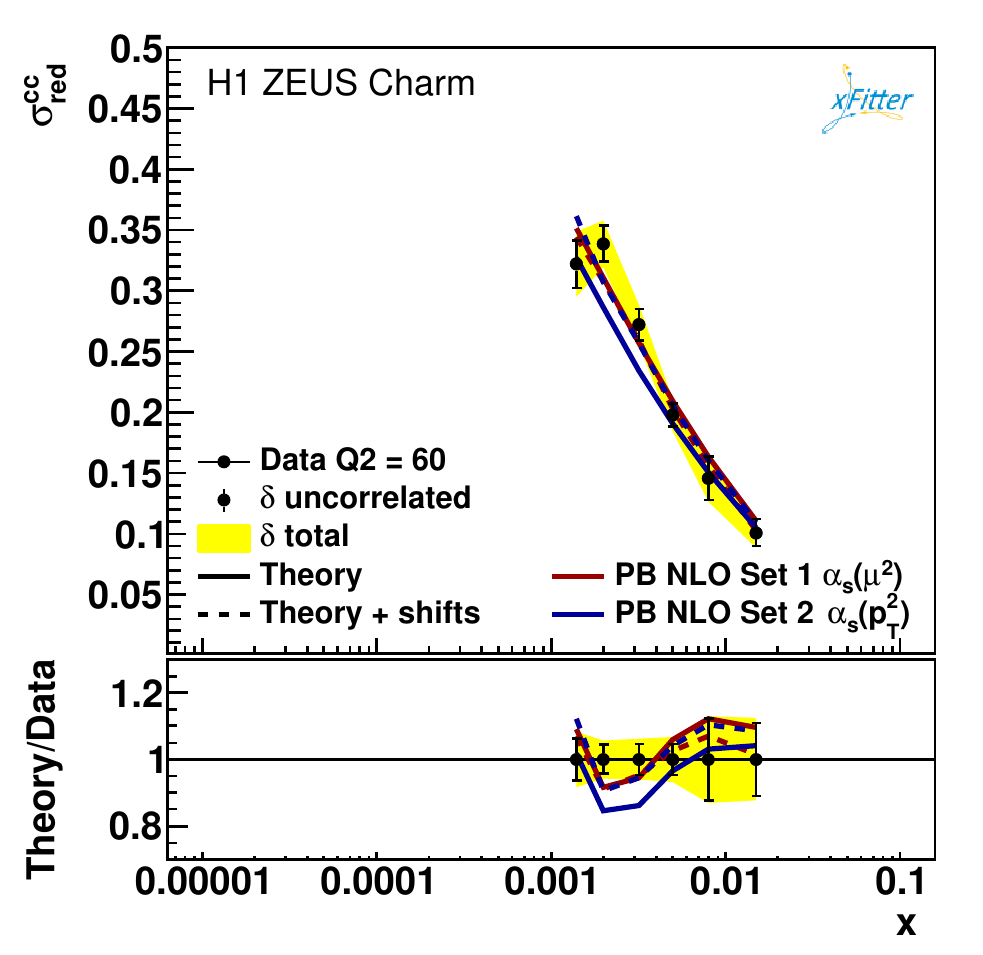}  
  \caption{\small Measurement of the reduced cross section obtained at HERA compared to predictions using Set~1 and Set~2. 
  Upper row: inclusive DIS cross section~\protect\cite{Abramowicz:2015mha}, lower row: inclusive charm production~\protect\cite{Abramowicz:1900rp}.
  The dashed lines include the systematic shifts in the theory prediction.}
\label{f2charm}
\end{center}
\end{figure}

In Fig.~\ref{f2charm} we show predictions for the inclusive DIS cross section and the inclusive charm cross section obtained from the two different parton distributions,  and compare them   with 
 the measurements from HERA~\cite{Abramowicz:2015mha,Abramowicz:1900rp}. 
While the inclusive DIS cross section is well described, the prediction using Set~2 differs from inclusive charm measurement at low $Q^2$ and small $x$. For values $x>0.001$ all predictions agree reasonably well with the data.
It has been checked explicitly that including the charm measurements in the fits does not significantly change the fit result (the charm data have too large an uncertainty compared to the precise inclusive measurements).
In Fig.~\ref{f2charm} the predictions including the systematic shifts are also shown, visually showing that the quality of the two different fits is similar.

\subsection{Transverse Momentum Dependent Parton Densities (TMD)} 
\label{sec:TMD}
Within the PB method both 
collinear and TMD densities can be determined, as the transverse momentum 
is calculated at every step of the branching process.
\begin{figure}[h!tb]
\begin{center} 
\includegraphics[width=0.48\textwidth]{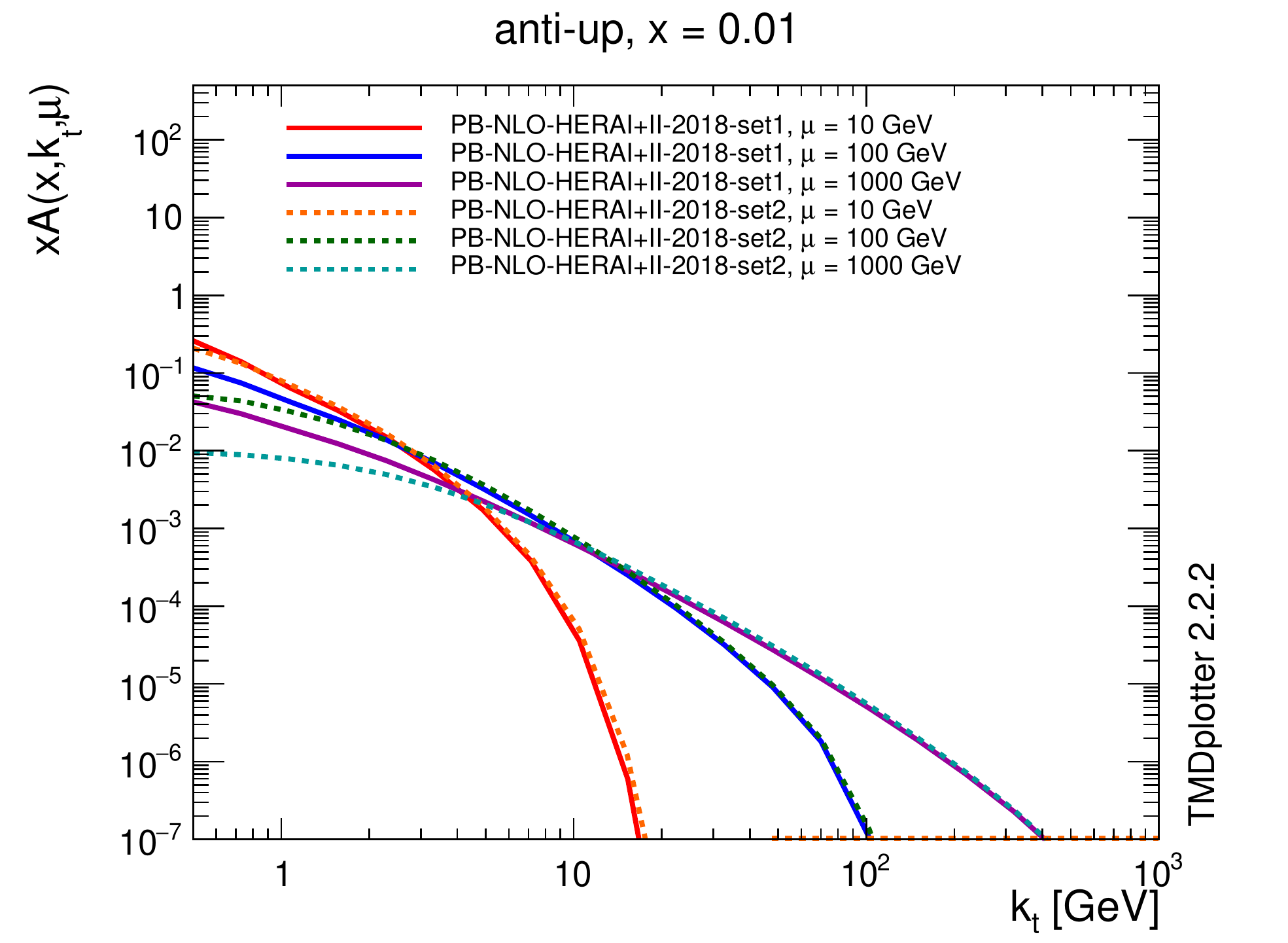} 
\includegraphics[width=0.48\textwidth]{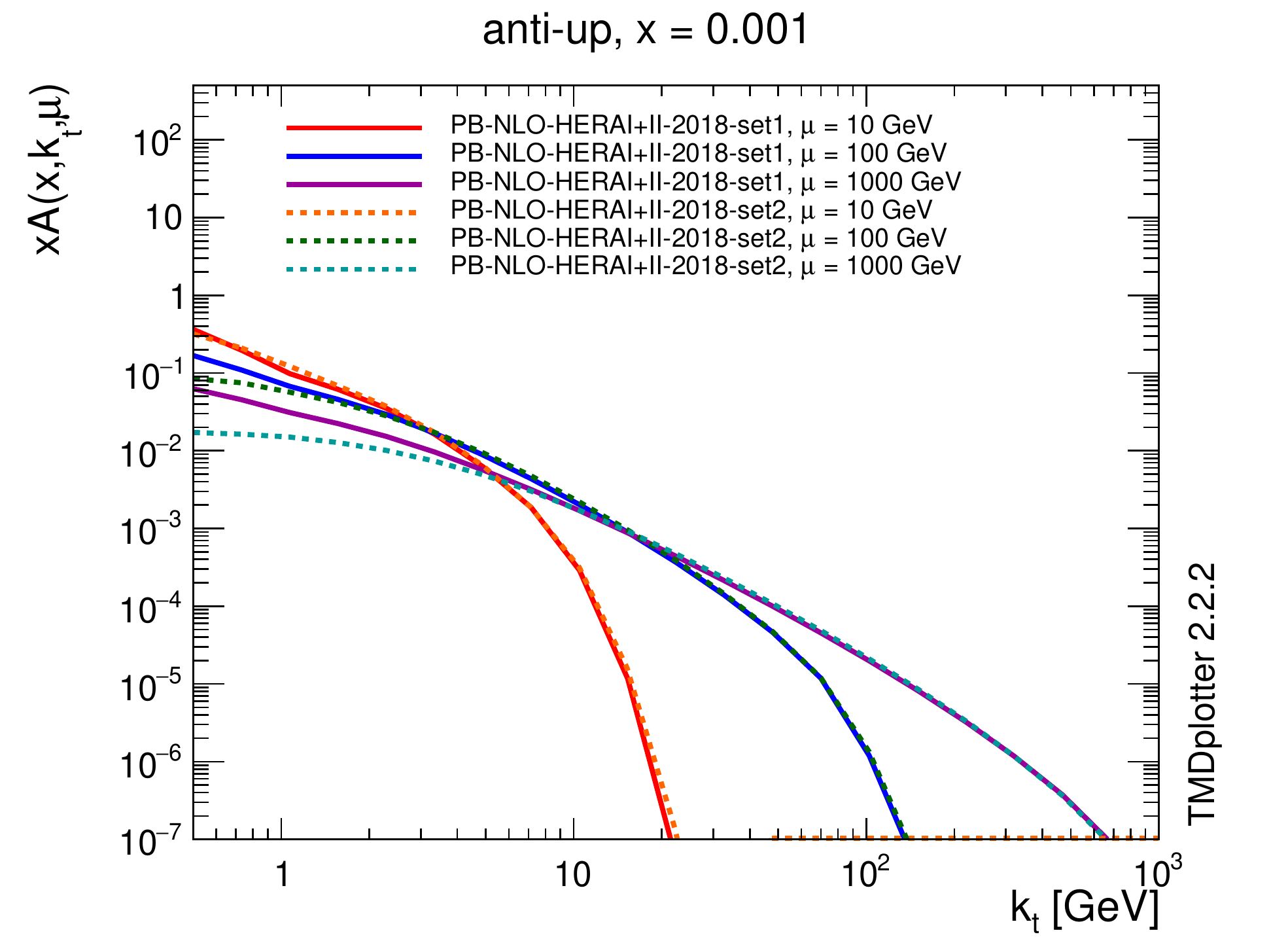} 
\includegraphics[width=0.48\textwidth]{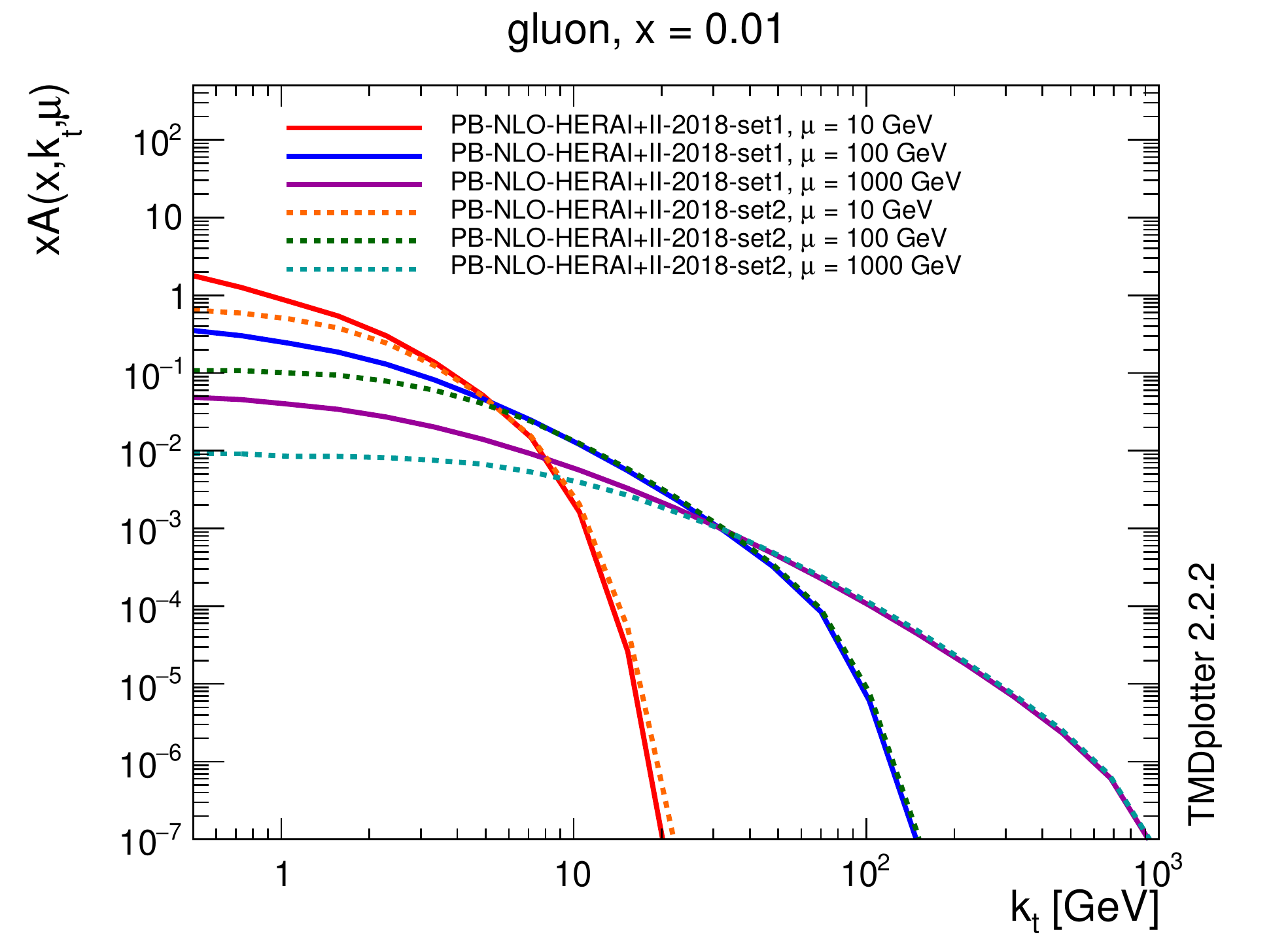} 
\includegraphics[width=0.48\textwidth]{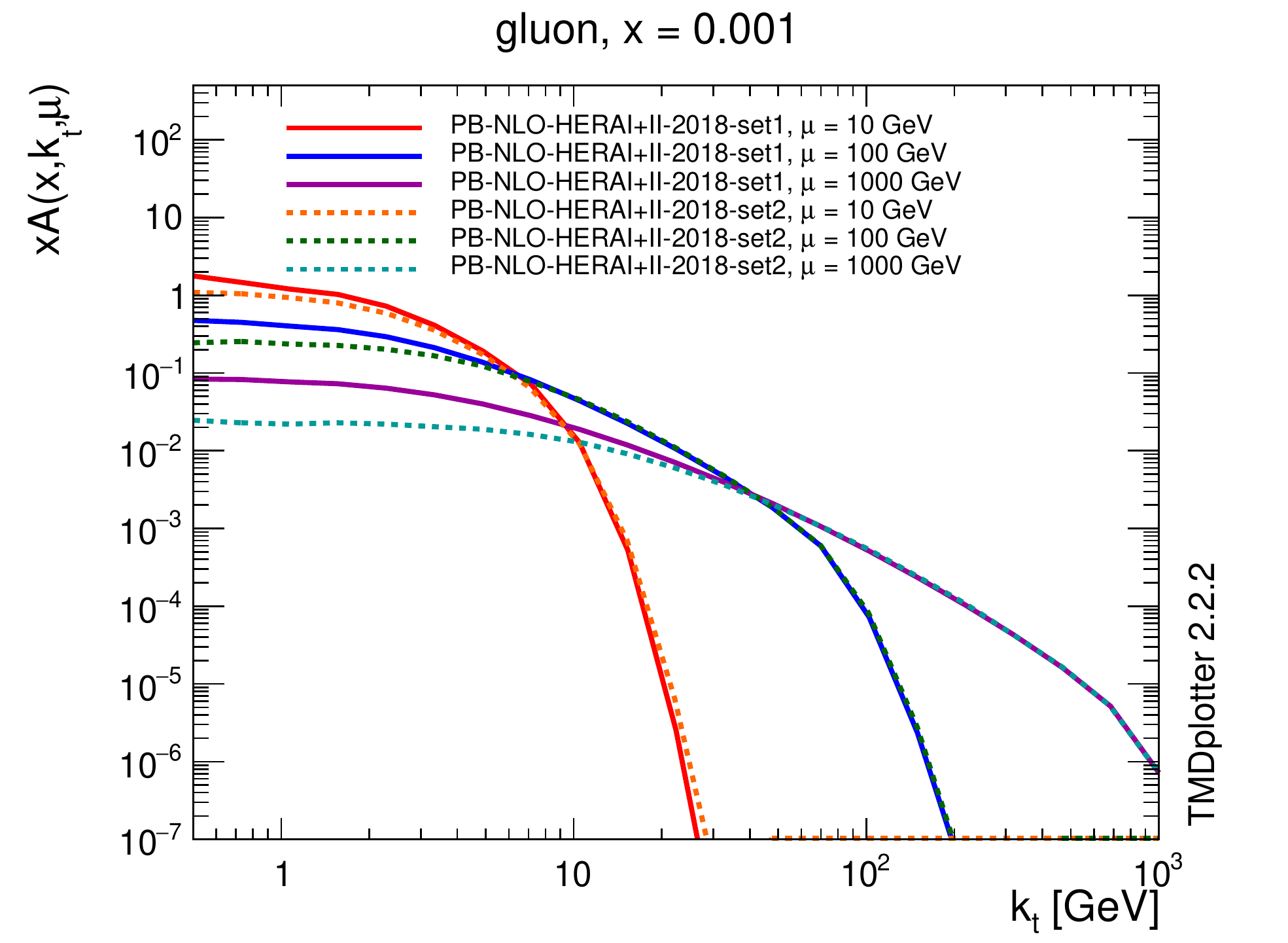} 
  \caption{\small Transverse Momentum Dependent parton densities (PB-NLO-2018-Set1 and PB-NLO-2018-Set 2) as a function of $\kt$ for different scales $\mu$.
  Upper row shows the densities for  $\bar{u}$, lower row the densities for gluons for two different values of $x$. 
  }
\label{TMD_pdfs1}
\end{center}
\end{figure} 
TMD parton densities can be obtained via the \PBM\ method once the relationship between 
kinematical variables and   
 evolution scale $\mu$ is specified, and
the transverse momentum at each individual branching is calculated with eq.~(\ref{ang-ordering}). 
The parameters for the starting distributions are obtained for the collinear parton densities by a fit to inclusive DIS cross section measurements, as described previously. 
The TMD  parton densities are then obtained from a convolution of the TMD kernel with the starting distribution as given in eq.~(\ref{TMD_kernel}). The 
starting distribution is taken from the collinear iTMD described in Sec.~\ref{sec:iTMD}.

 \begin{tolerant}{300}
In Fig.~\ref{TMD_pdfs1} we show the TMD parton densities for $\bar u$-quarks and gluons as a function of the  transverse momentum $k_t =  \sqrt{{\bf k }^2  }$ for different values of the evolution scale $\mu= 10,\, 100,\, 1000$~GeV and different values of $x$ for Set 1 and  Set 2. One can clearly see  that both sets give identical results for larger $\kt$, while they are different for small $\kt$, a consequence of the different scale choices for the argument of \alphas .
\end{tolerant}

\begin{figure}[htb]
\begin{center} 
\includegraphics[width=0.32\textwidth]{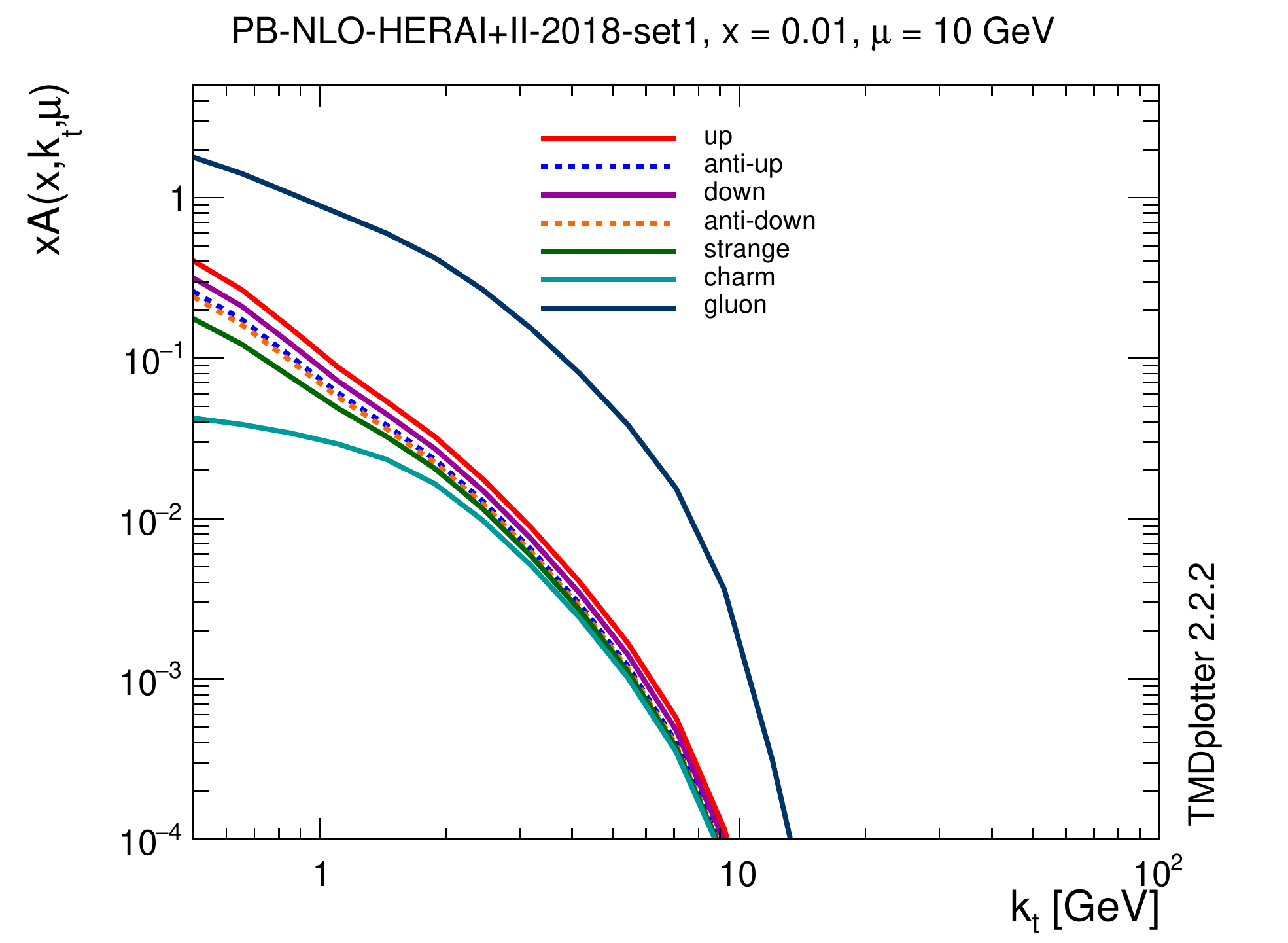} 
\includegraphics[width=0.32\textwidth]{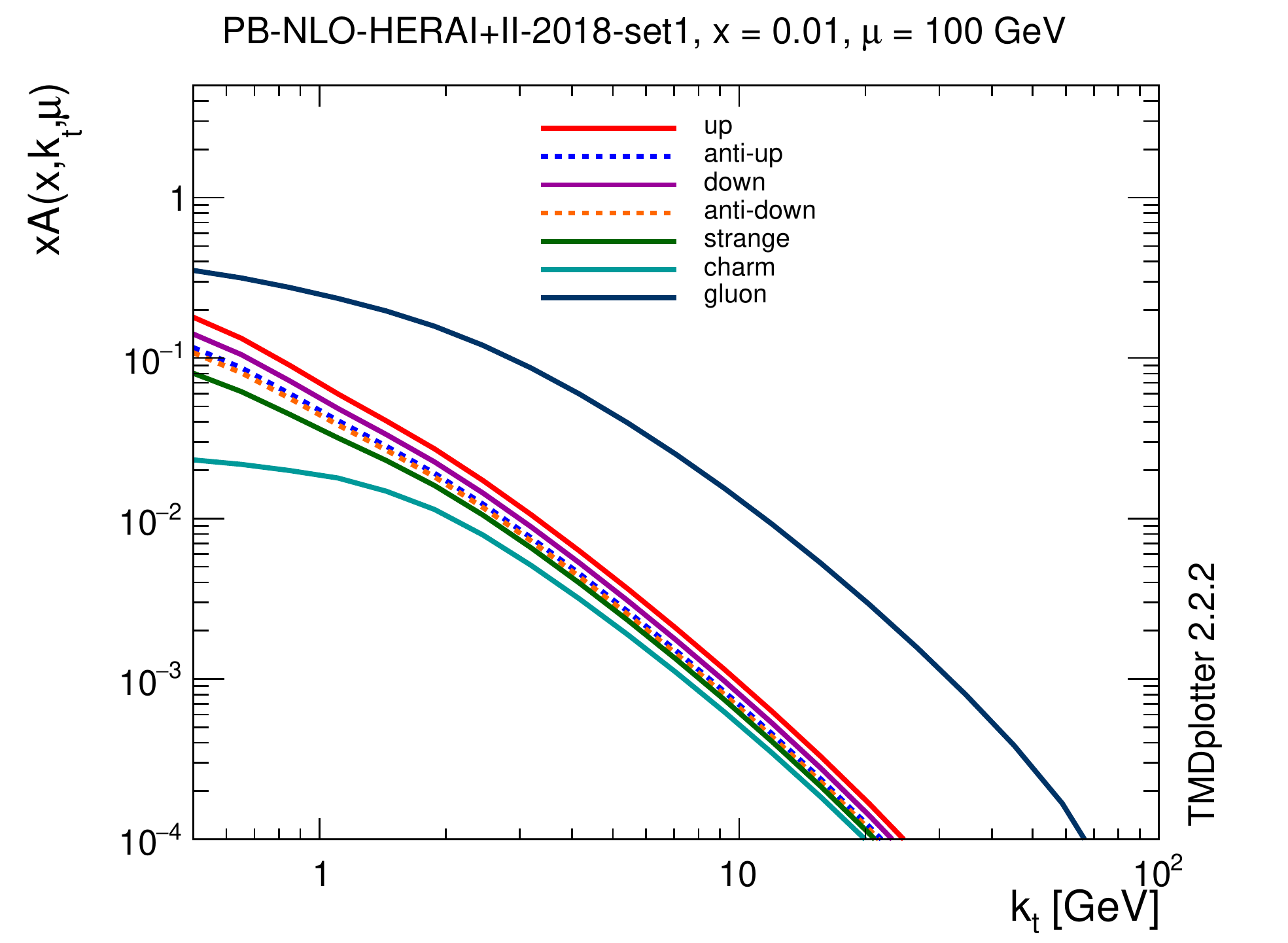} 
\includegraphics[width=0.32\textwidth]{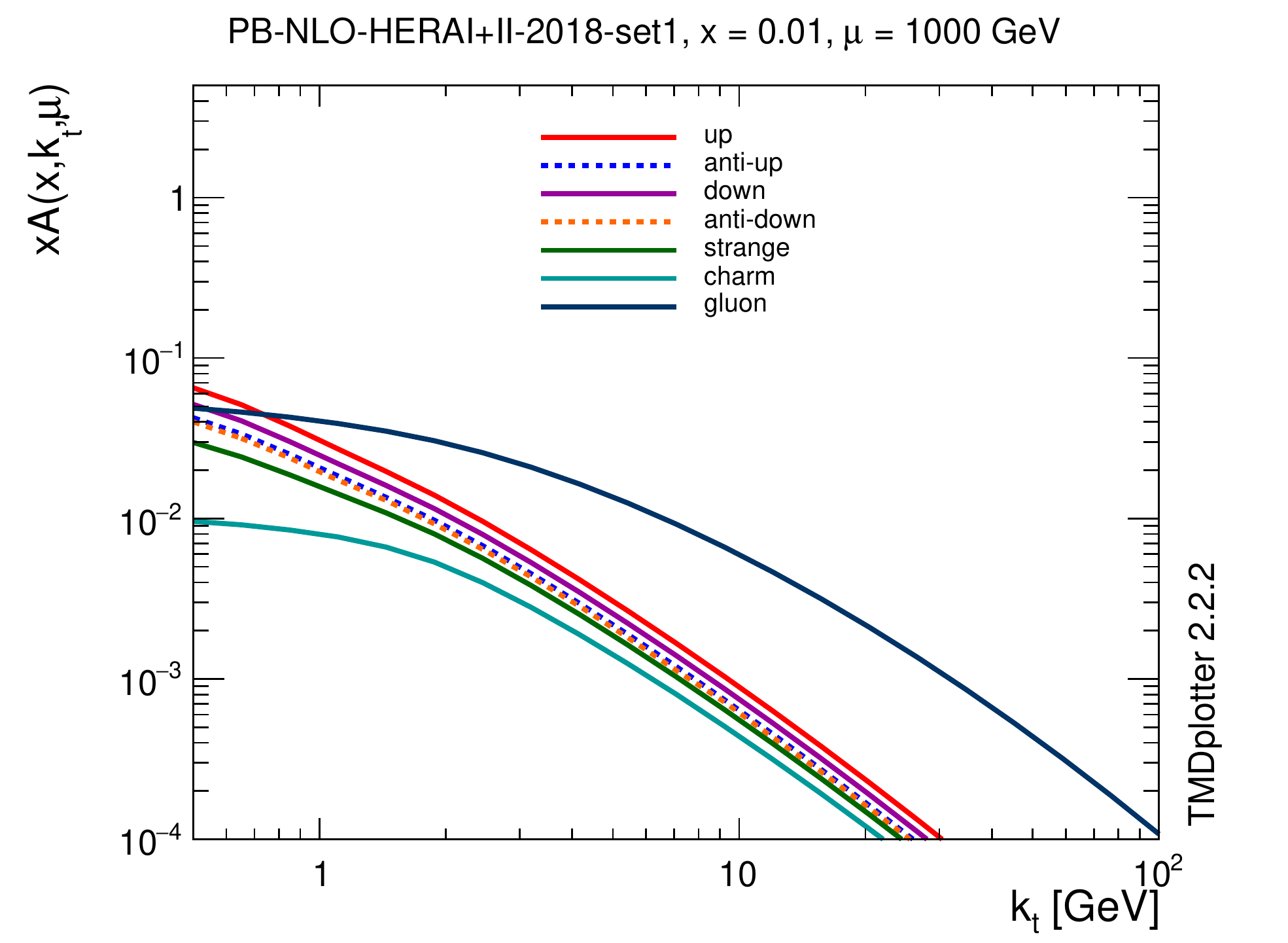} 
\includegraphics[width=0.32\textwidth]{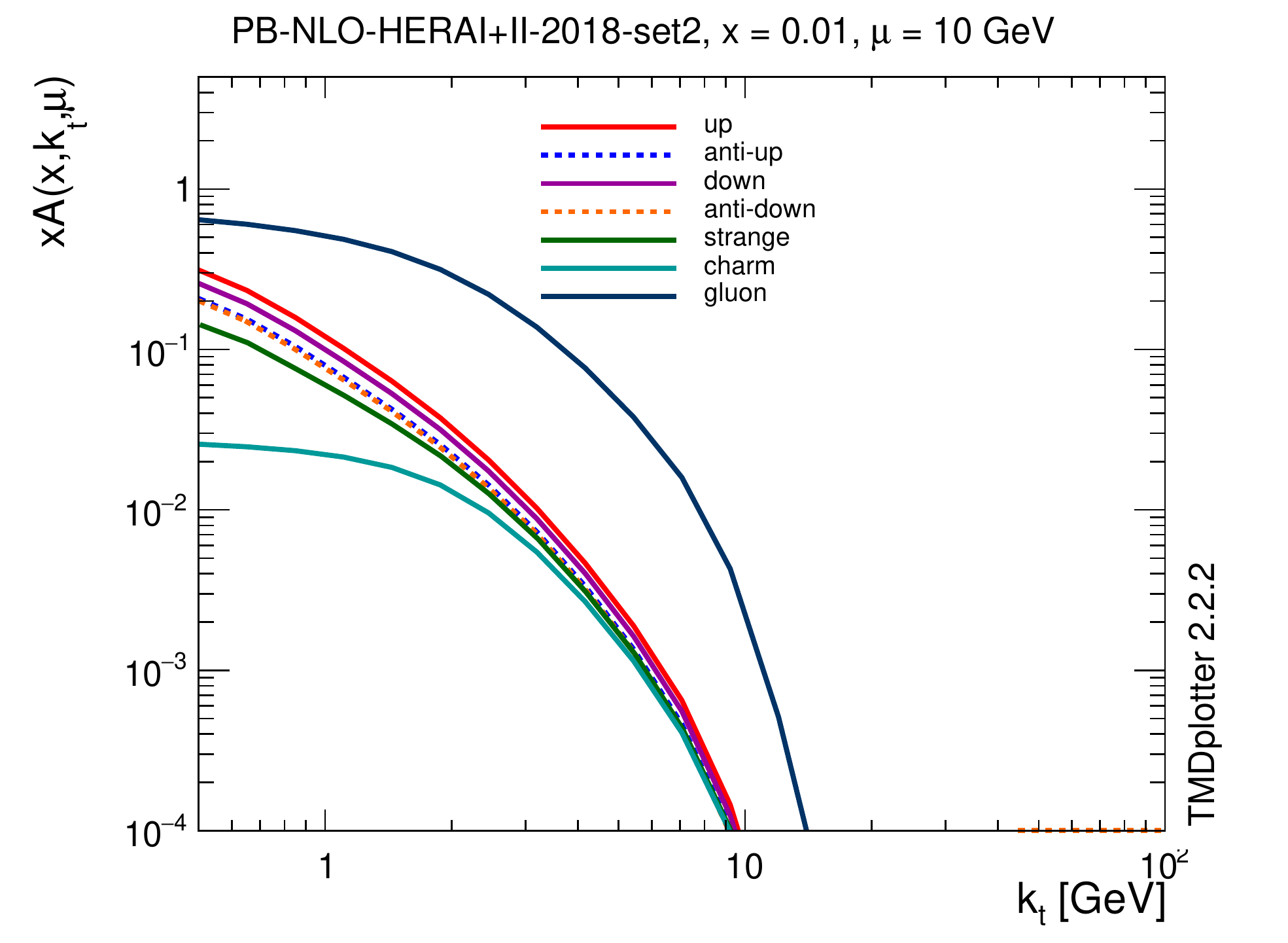} 
\includegraphics[width=0.32\textwidth]{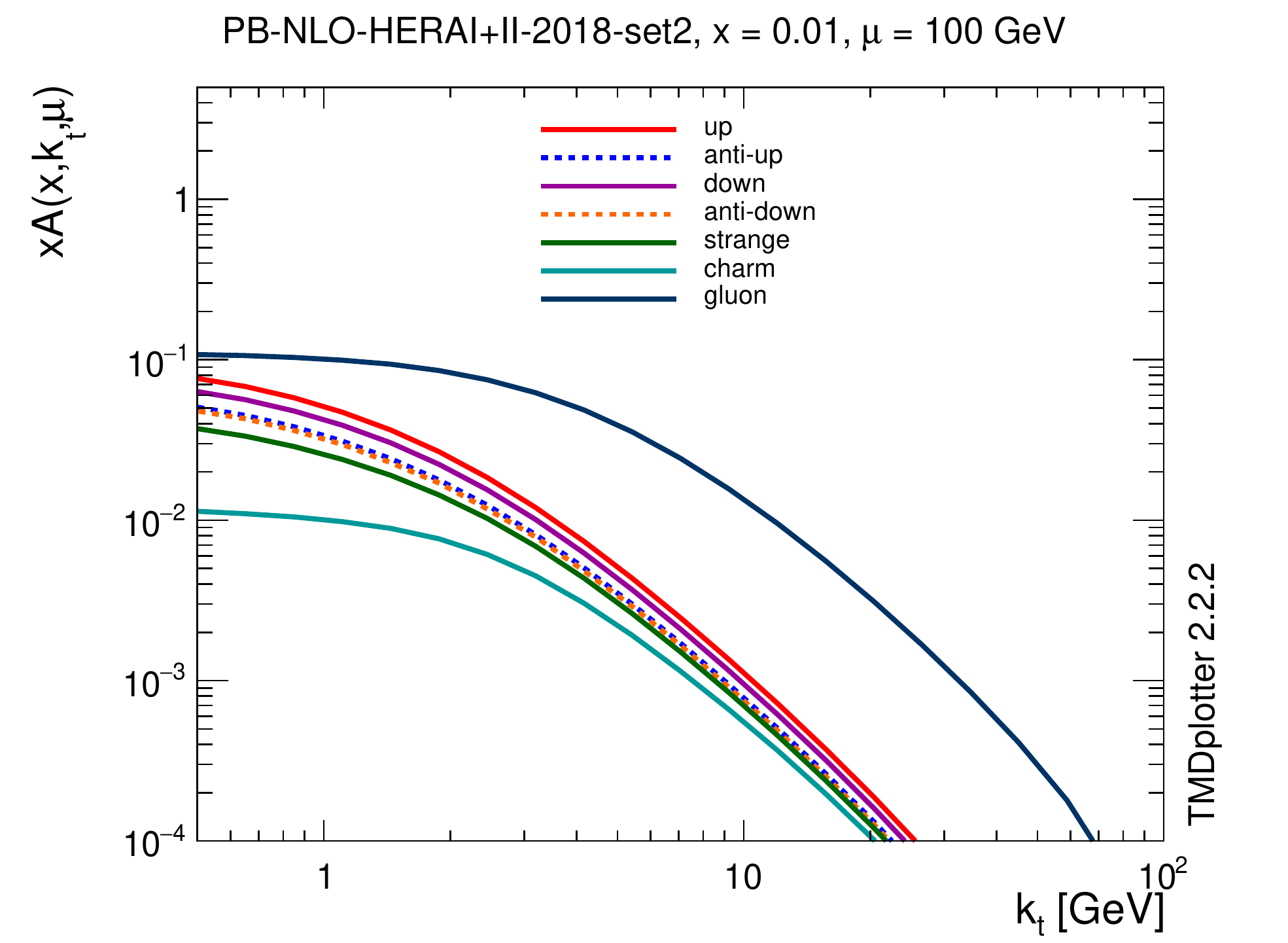} 
\includegraphics[width=0.32\textwidth]{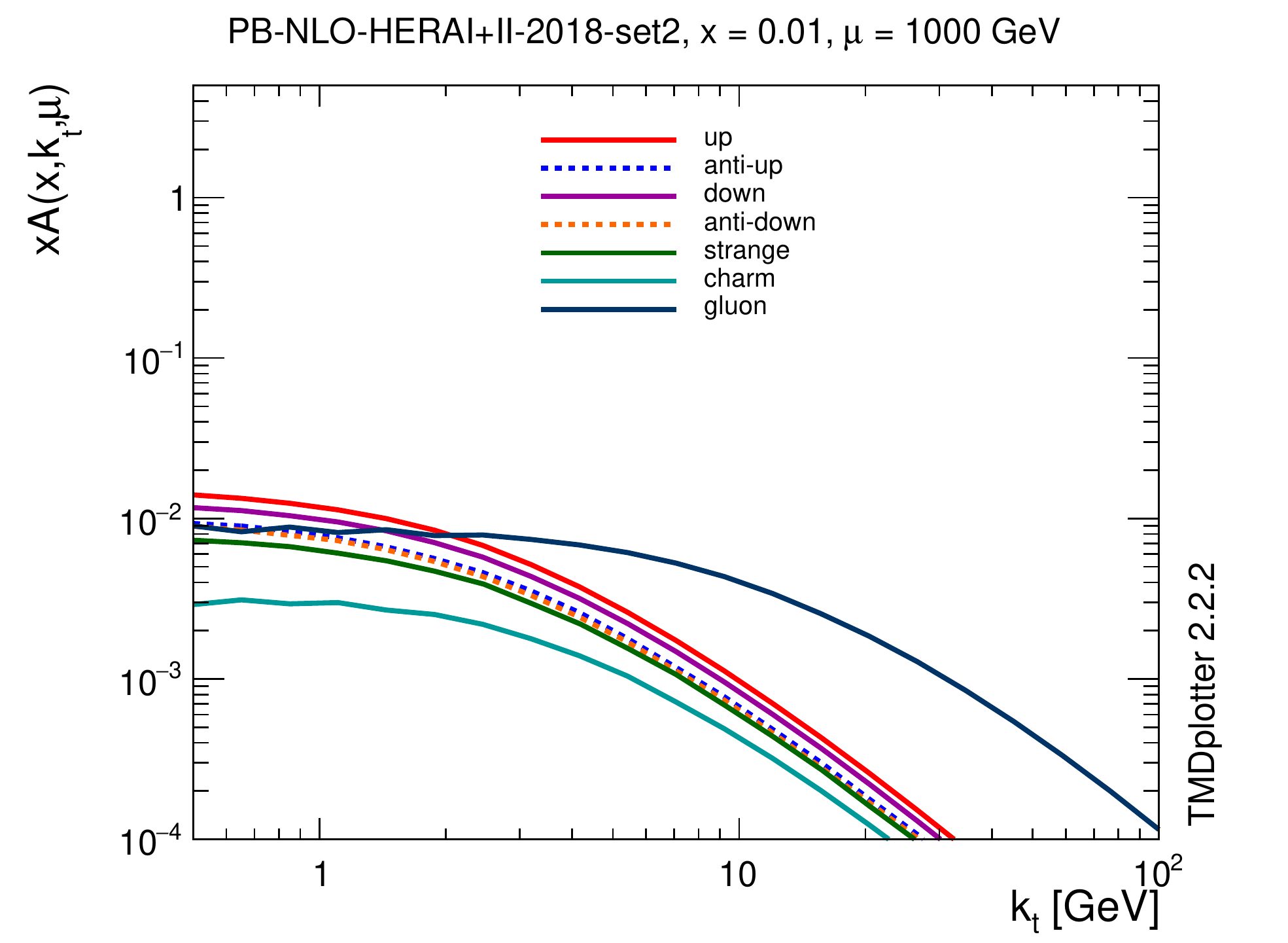} 
  \caption{\small Transverse Momentum Dependent parton densities (PB-NLO-2018-Set1 upper row and PB-NLO-2018-Set2 lower row) as a function of $\kt$ for different scales $\mu$ at $x=0.01$ for all flavors.
  }
\label{TMD_pdfs2}
\end{center}
\end{figure}

In Fig.~\ref{TMD_pdfs2} the parton densities for all flavors are shown as a function of $\kt$ at $x=0.01$ and for different values of the evolution scale $\mu = 10,\, 100,\, 1000$ GeV. The large scales are relevant for phenomenology at the LHC, and it is interesting to observe that the transverse momenta extend to very large values, up to the values of the factorization scales (for $\mu = 1$~TeV the transverse momenta extend to $\kt \sim 1$~TeV).  However, the large $\kt$ values are suppressed compared to smaller ones. 
The different quark flavors show a different behavior at small $\kt$,  coming essentially from the no-branching probability times the starting distribution  (first term in eq.~(\ref{evoleqforA})), while they are very similar at larger $\kt$, a result of perturbative splittings (second term in eq.~(\ref{evoleqforA})).

\begin{figure}[htb]
\begin{center} 
\includegraphics[width=0.495\textwidth]{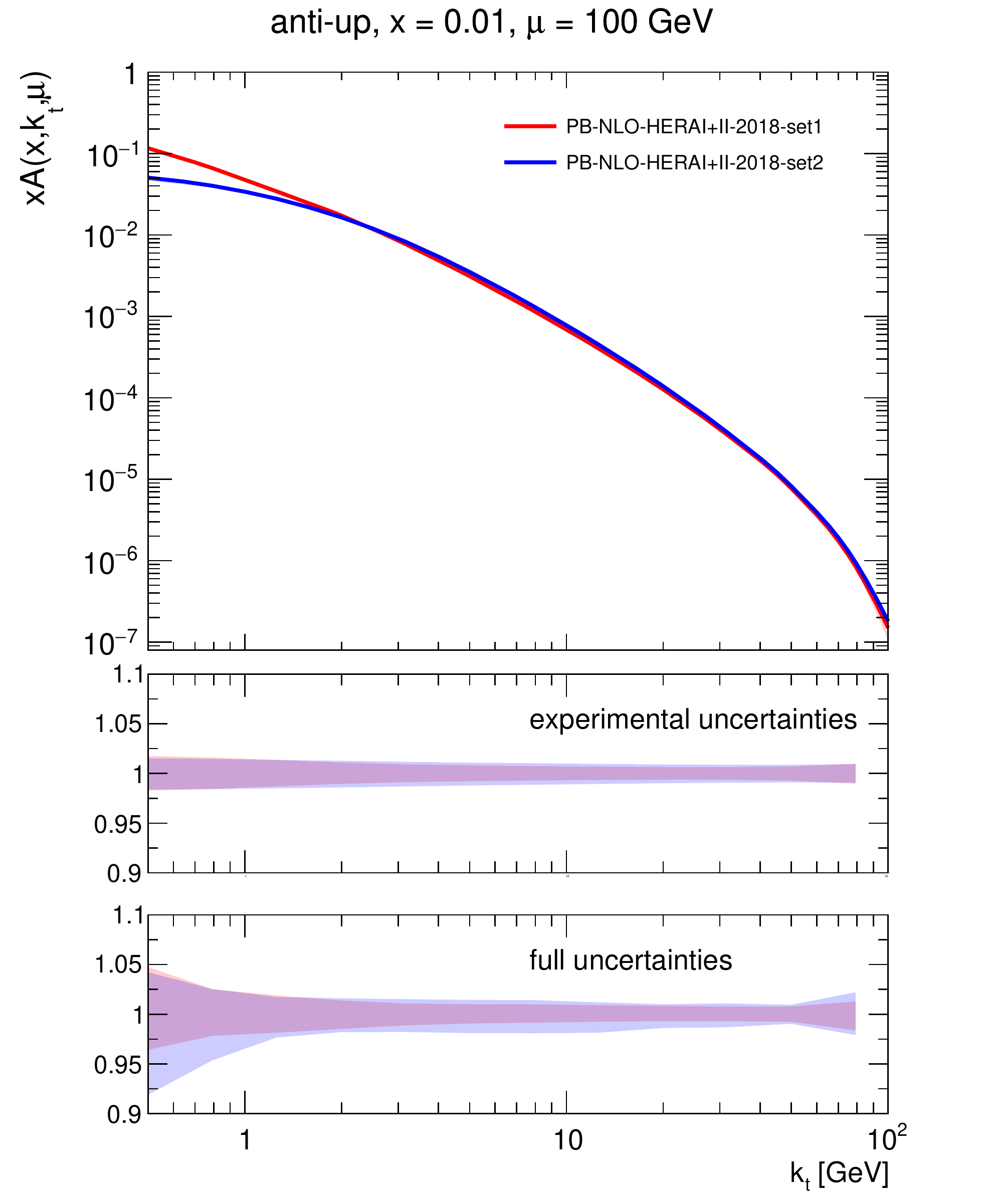}
\includegraphics[width=0.495\textwidth]{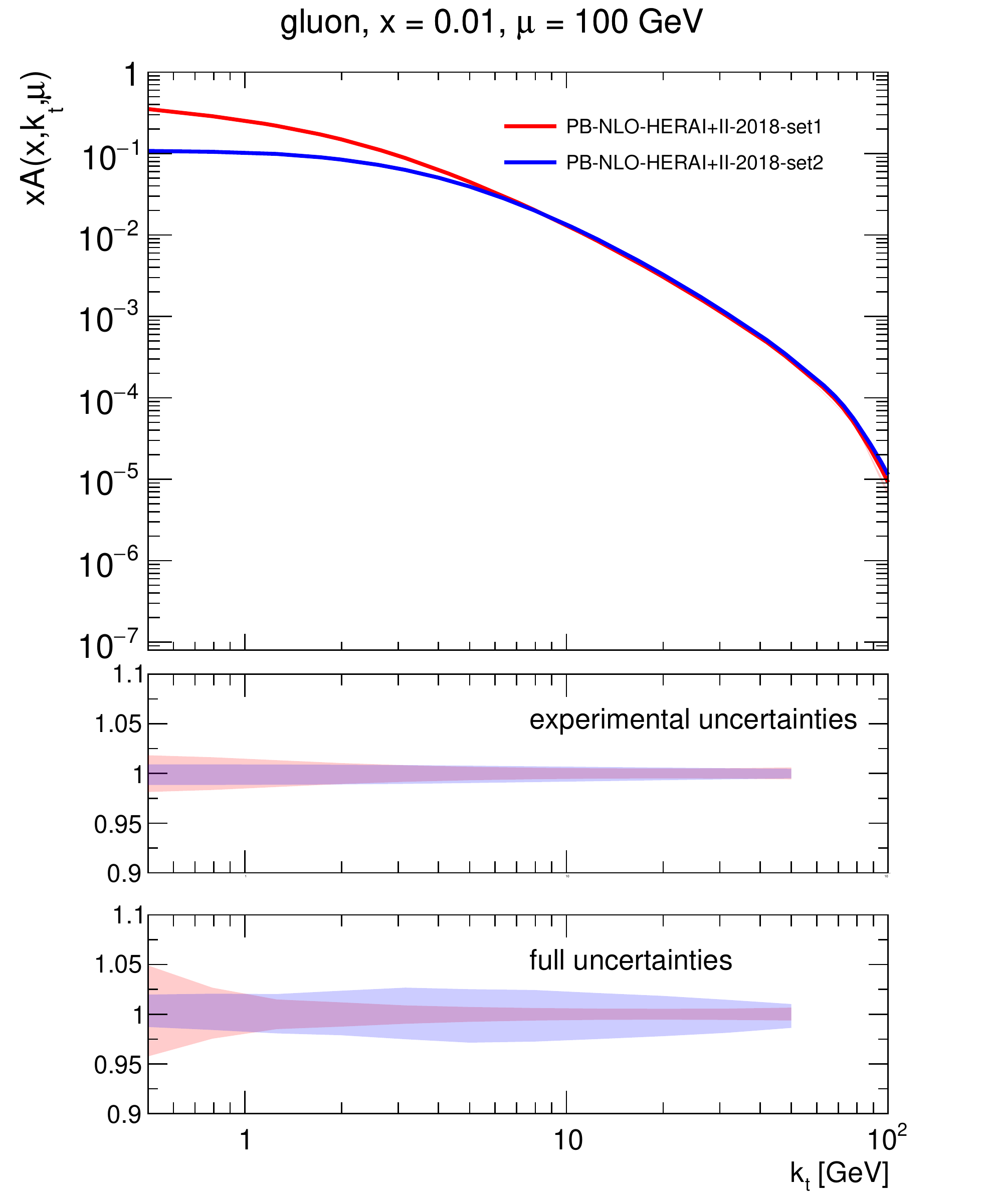}
  \caption{\small Transverse Momentum Dependent parton densities for $\bar u$ and gluon  from Set~1 and Set~2 as a function of $\kt$ for  $\mu=100$~GeV at $x=0.01$. In the lower panels 
we show the relative uncertainties coming  from experimental uncertainties as well as the total of experimental and model uncertainties.
  }
\label{TMD_pdfs3}
\end{center}
\end{figure}

In Fig.~\ref{TMD_pdfs3} the gluon and $\bar{u}$ densities as a function of the transverse momentum are shown for $\mu=100$~GeV and  $x=0.01$ together with the uncertainty bands obtained from the fits. The panels show the uncertainties coming from the experimental sources as well as the total uncertainty coming from experimental and model sources separately.
Although only collinear splitting functions are used, and the fit was obtained with collinear parton densities, a $\kt$ dependence of the uncertainties is obtained. At small $\kt$ essentially the first term in eq.~(\ref{evoleqforA}) contributes  without any resolvable branching and the uncertainty comes from the starting distribution at $x$, while at large $\kt$ several branching may have occurred and therefore the uncertainty comes from the starting distribution at $x/z\gg x$.
The experimental uncertainties are small over the whole range, while the model dependent uncertainties dominate. 

\begin{figure}[htb]\begin{center} 
\includegraphics[width=0.495\textwidth]{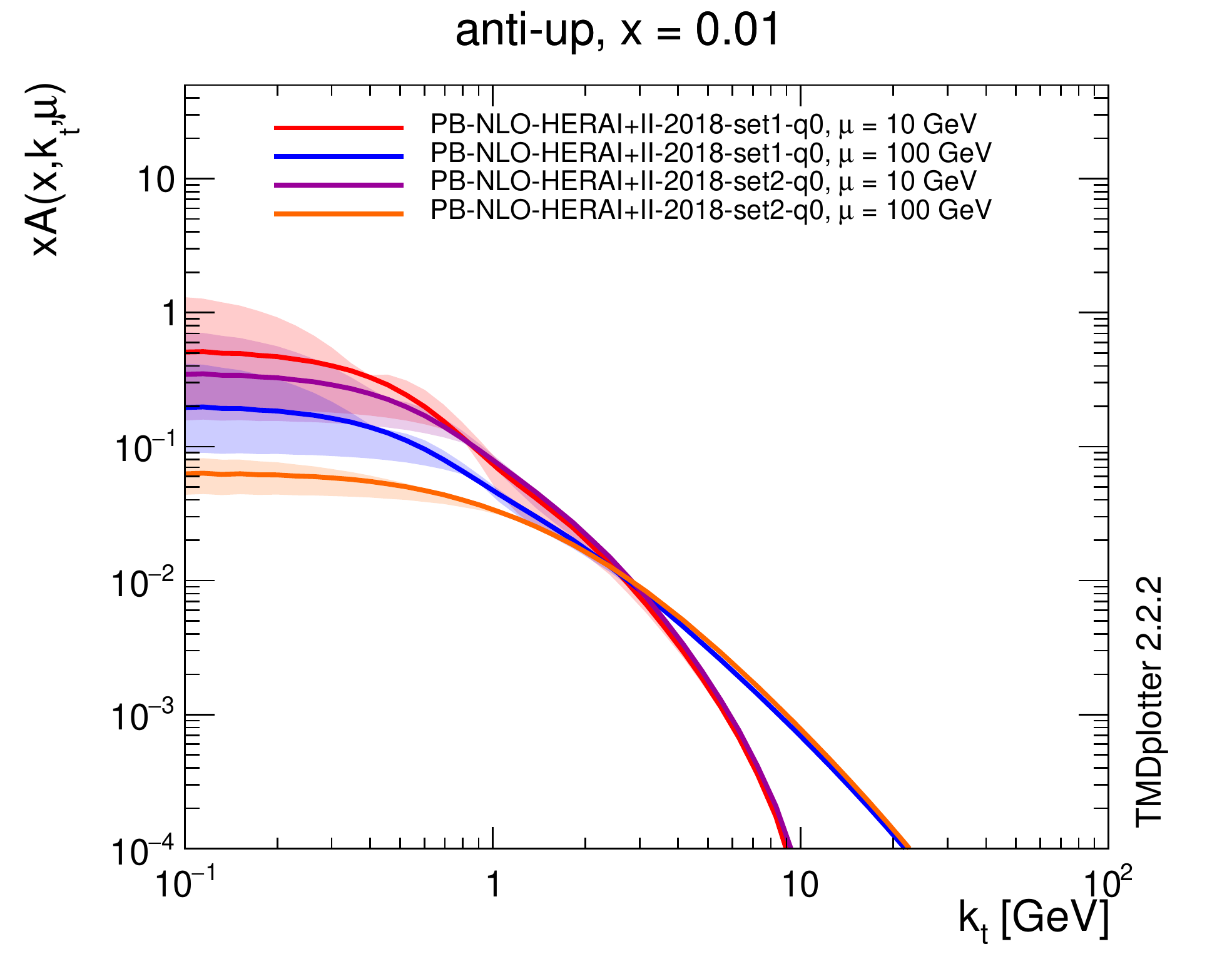}
\includegraphics[width=0.495\textwidth]{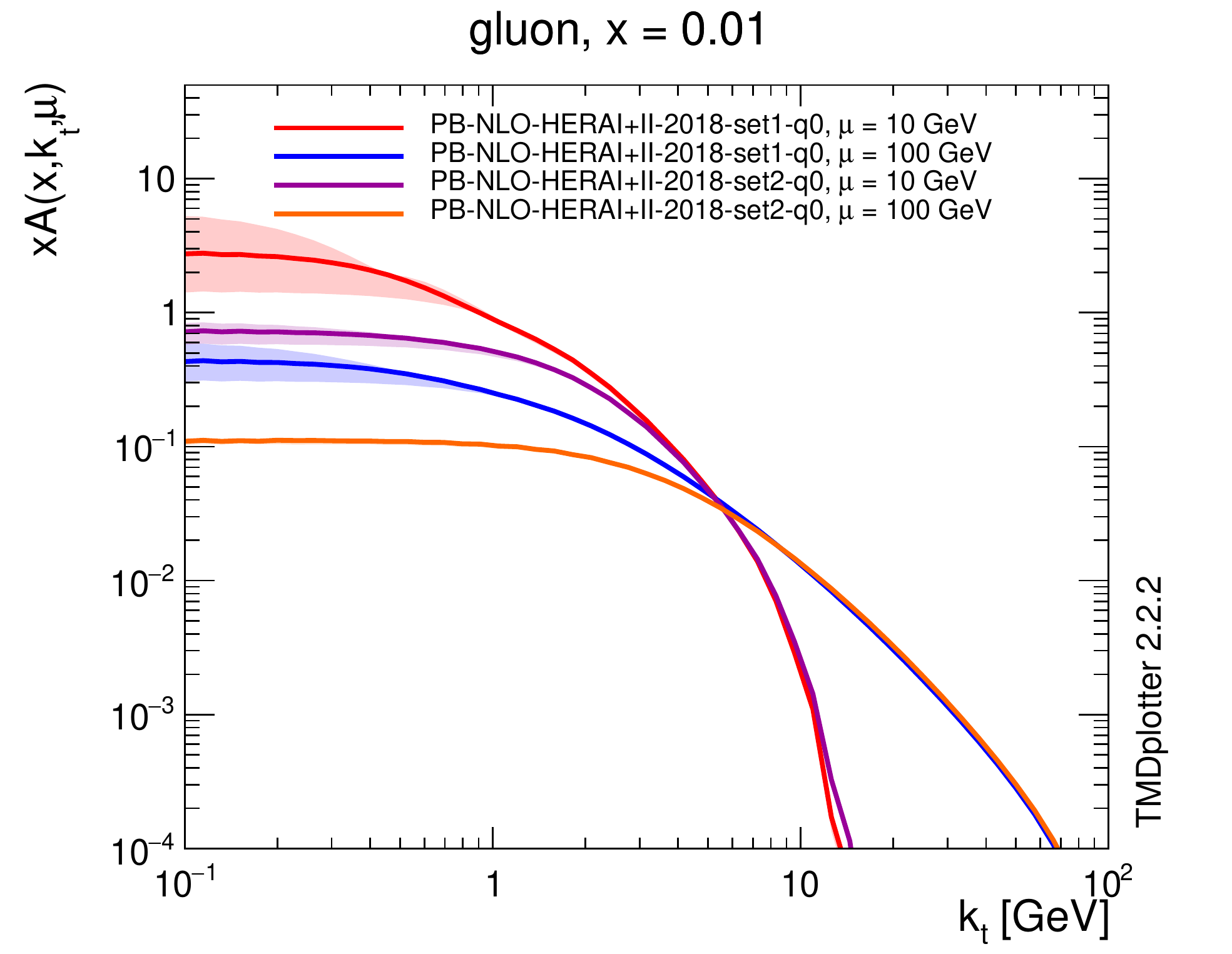}
  \caption{\small Transverse Momentum Dependent parton densities ($\bar u$ and gluon) from Set~1 and Set~2 as a function of $\kt$ for  $\mu=10 (100)$~GeV at $x=0.01$, when the width of the intrinsic transverse momentum distribution is  varied by a factor of two.}
\label{TMD_q0}
\end{center}
\end{figure}

The parametrization of the intrinsic transverse momentum distribution is another uncertainty. With the fit to inclusive DIS data, this distribution cannot be further constrained. In Fig.~\ref{TMD_q0} we show the TMD distribution for gluon and $\bar{u}$ for Set~1 and Set~2 at  $\mu=10 (100)$~GeV and $x=0.01$ when $q_0$ in $ {\cal B}(\ktz^2,\mu_0^2)$ is varied from $q_0=0.25$ ~GeV to $q_0=1$~GeV. We do not include the variation of $q_0$ as a systematic uncertainty, since it is not constrained by the fit (in future we plan to use also $Z$-boson transverse momentum spectra, which would constrain $q_0$). 

The resulting TMD parton densities, PB-NLO-2018-Set1  and PB-NLO-2018-Set2, including uncertainties (as well as with variation of $q_0$) are available in \TMDlib \cite{Hautmann:2014kza}. 
The  \TMDplotter 
\cite{TMDplotter2,Connor:2016bmt} interface allows easy and fast comparison to other TMDs, once they are made publicly accessible and available in TMDlib.

\section{Application to \boldmath$Z$-boson  production at the LHC}

The transverse momentum spectrum of $Z$ bosons in Drell-Yan (DY) production at small values of transverse momentum $q_T$ cannot be described by fixed-order perturbative calculations, and resummation of soft gluon emissions to all orders in $\alphas$ is needed. See e.g.~\cite{Angeles-Martinez:2015sea} for a recent discussion.   The DY $q_T$ spectrum can be described by the CSS method~\cite{Collins:1984kg,Landry:2002ix,Collins:2012ss,Catani:2015vma}  using TMD factorization at small $q_T$~\cite{Bacchetta:2017gcc,Scimemi:2017etj}, or by  parton showers within Monte Carlo event generators~\cite{Hoche:2017hno}.  
The ATLAS and CMS experiments at the LHC have measured the $q_T$ spectrum of the  $Z$-boson \cite{Aad:2011gj,Chatrchyan:2011wt,Aad:2015auj}.

The TMD distributions obtained from HERA DIS measurements can be used to predict the DY $q_T$ spectrum of the  $Z$-boson at LHC energies. Since we are interested in the low-$q_T$ region, we use the LO expression for  $Z$ production matrix elements.\footnote{In practical terms we use an LHE (Les-Houches Event) file~\cite{Alwall:2006yp} for $q\bar{q}\to Z $ obtained from the \pythia\ event generator~\cite{Sjostrand:2014zea} with on-shell initial partons.} The transverse momentum of the initial state partons is calculated according to the TMDs and added to the event record in such a way that the mass of the produced DY pair is conserved, while the longitudinal momenta are changed accordingly. This procedure is common in standard parton shower approaches~\cite{Bengtsson:1986gz,Sjostrand:2014zea} and is implemented in the \cascade\ package~\cite{Jung:2010si,Jung:2001hx} (version newer than \verb+2.4.X+) where events in HEPMC~\cite{Dobbs:2001ck} format are produced, for further processing with Rivet \cite{Buckley:2010ar}. The importance of the proper inclusion of transverse momentum effects from parton showers has been pointed out in Ref.~\cite{Dooling:2012uw,Hautmann:2012dw}. With the TMD distributions described here, these effects can be included already at the level of the cross section calculation.

\begin{figure}[htb]
\begin{center} 
\includegraphics[width=0.495\textwidth]{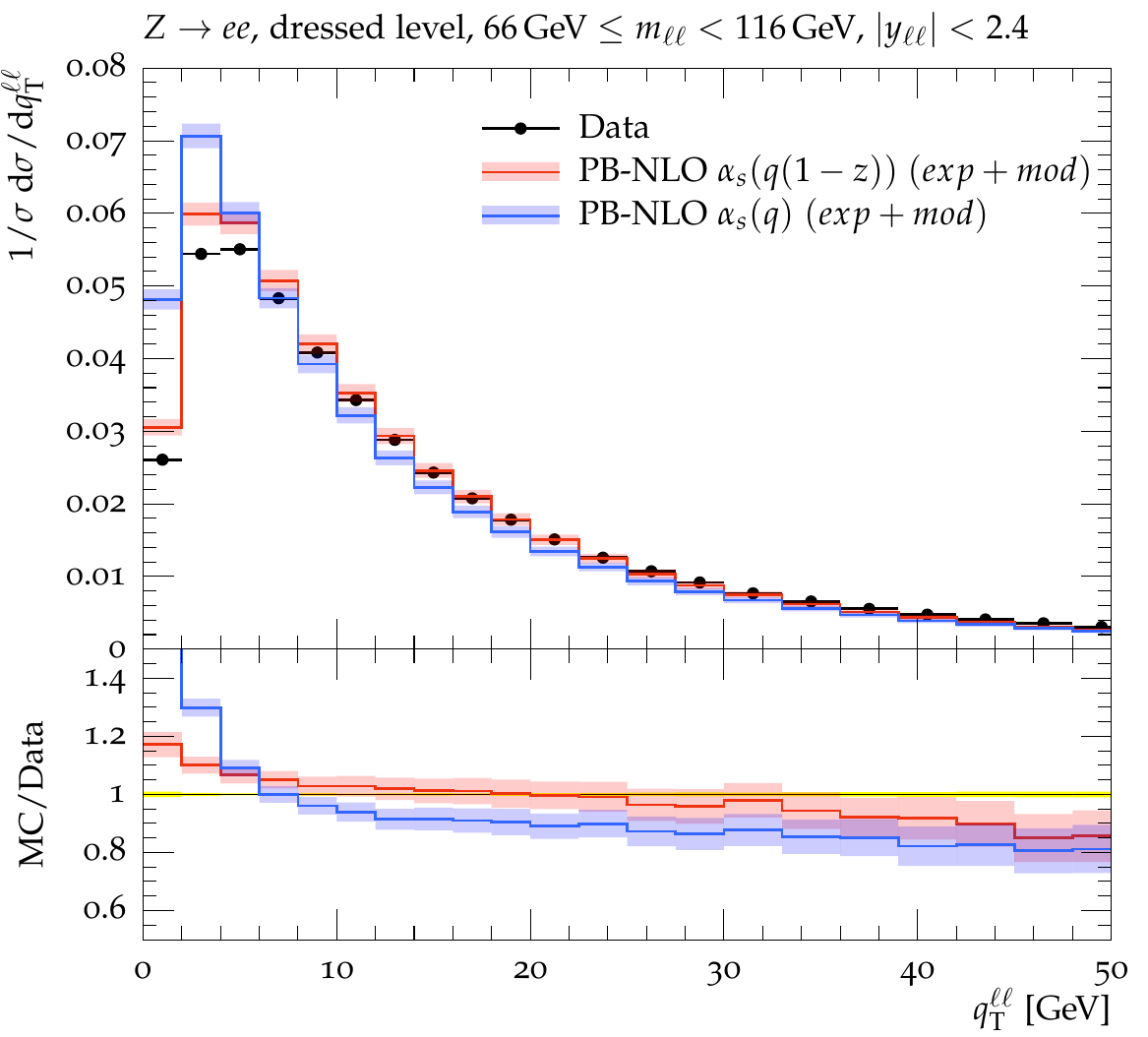} 
\includegraphics[width=0.495\textwidth]{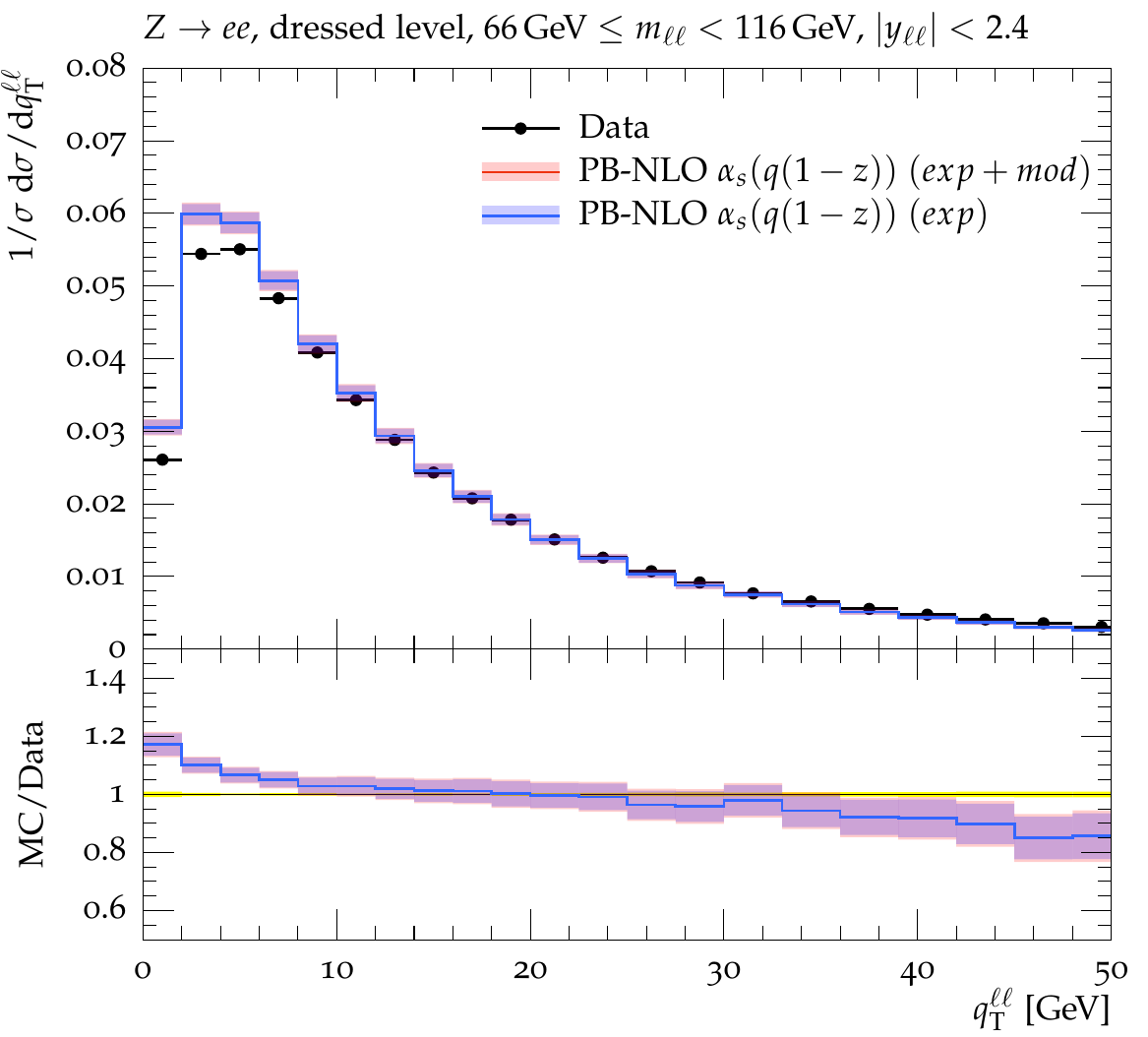} 

\caption{\small Transverse momentum $q_T$ spectrum of $Z$-bosons obtained from the two TMDs,  compared with measurements from \protect\cite{Aad:2015auj}. Left: comparison of predictions using Set~1 and Set~2 including the full (experimental and model) uncertainties. Right: prediction using Set~2, with experimental and full uncertainties separated (the difference is very small).
  }
\label{Zpt}
\end{center}
\end{figure}

In Fig.~\ref{Zpt} (left) we show the predictions for the transverse momentum spectrum of the $Z$-boson obtained with the two TMD distributions, compared with the measurements of ATLAS~\cite{Aad:2015auj}. The uncertainties coming from experimental and model sources are shown for both Set~1 and Set~2 with the colored bands (Fig.~\ref{Zpt} left); the experimental and full uncertainties are shown for Set~2  in Fig.~\ref{Zpt} (right). The difference between the full and experimental uncertainties from the fit is very small.  

In general the shape of the spectrum is described by both TMD fits. The TMD Set~2, applying the transverse momentum as the renormalisation scale (instead of the evolution scale $\mu$), provides a significantly better description of the transverse momentum spectrum of the $Z$-boson, coming from the different $\kt$ spectrum of the TMD already visible in Fig.~\ref{TMD_pdfs1}.
One should note that no adjustment of any parameter is made, and that the TMDs are entirely constrained 
by the fits to inclusive DIS data. The description of the transverse momentum spectrum of the $Z$-boson obtained with the PB-TMD Set~2 is of similar quality as the NLO+NNLL prediction of Ref.~\cite{Bizon:2018foh}, however, one should note, that the approach of PB-TMDs is more general and can be applied directly to other processes as well without further modification.

\section{Conclusion}

The parton branching method has been used to determine a first complete set of collinear and TMD parton densities from fits to precision DIS data over a large range in $x$ and $Q^2$ as measured at HERA.
The parton densities are obtained with NLO DGLAP splitting functions and 2-loop \alphas\ with $\alphas (M_Z) = 0.118$. The renormalisation scale in the evolution has been chosen to be the evolution scale $\mu_i$ (Set 1) or the transverse momentum $q_{t\,i}$ (Set 2). Two different collinear and TMD sets are obtained for these different choices, both giving a similar $\chi^2/ndf = 1.2 $. The obtained parton densities are valid over a wide range in $x$ and scale $\mu$, up to the multi-TeV scale,  relevant for LHC physics.

Experimental uncertainties of the fit are obtained using the Hessian method with $\Delta \chi^2=1$ and model dependent uncertainties are determined. 

The obtained TMDs are applied to calculate the transverse momentum spectrum of the $Z$-boson in 
DY production at LHC energies. Good agreement with the measurement is observed if angular ordering is applied. The uncertainties of the prediction come only from the TMD uncertainties determined in the fit to HERA measurements.

For the first time, precision DIS measurements have been used to obtain both collinear and TMD parton densities, including uncertainties, over a wide range in $x$ and $\mu$ values, which are relevant for LHC and future collider phenomenology as well as for low-energy and small-$\kt$ physics.

\noindent

{\bf Acknowledgments.}
We are grateful for many discussions with the xFitter developers team, in particular with R. Placakyte and A. Glazov.
FH acknowledges the support and hospitality of the CERN Theory Division, of  DESY, Hamburg, while part of this work was being done. HJU thanks the Polish Science and Humboldt Foundations for the Humboldt Research fellowship during which part of this work was completed.

\section*{Appendix}

\begin{figure}[htb]
\begin{center} 
\includegraphics[width=0.495\textwidth]{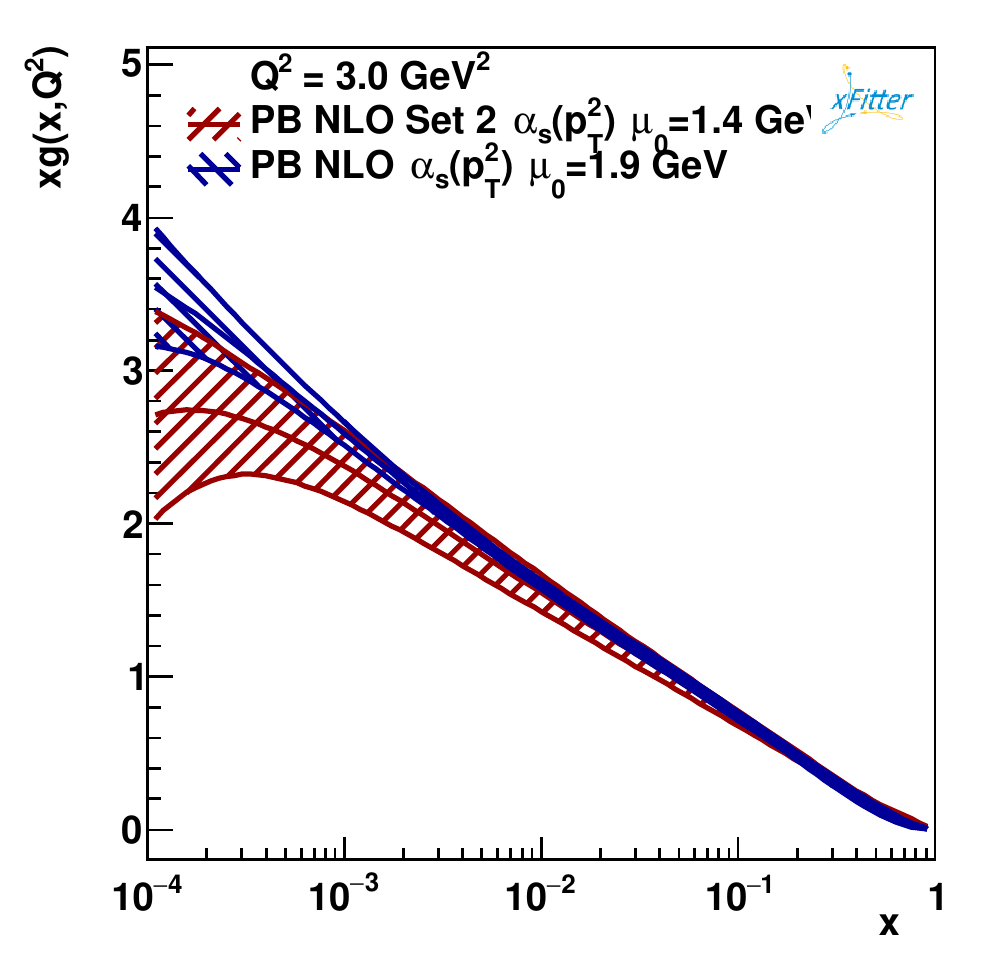} 
\includegraphics[width=0.495\textwidth]{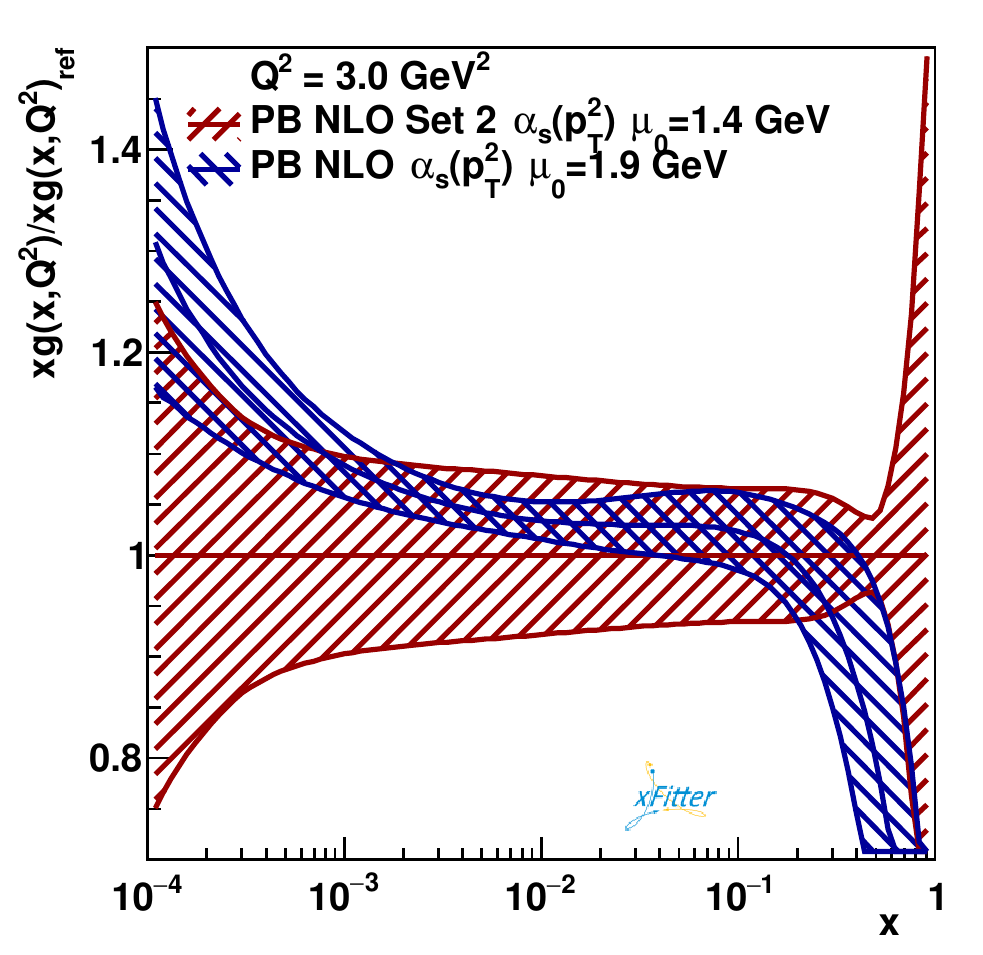} 

\caption{\small Comparison of gluon density of Set~2 type obtained at $\mu_0^2 =1.4$~GeV$^2$ and $\mu_0^2 =1.9$~GeV$^2$ at a scale of $Q^2 =3$~GeV$^2$. The ratio of the gluon densities is shown with respect to the default Set~2.  The uncertainties for the new fit include only those from experimental sources, the uncertainties for Set~2 are the same as in Fig~\ref{collinear_pdfs_ratio}.
 }
\label{mu0variation}
\end{center}
\end{figure}

In Fig.~\ref{mu0variation} we show  a comparison of the gluon  density of Set~2 ($\mu_0^2 =1.4$~GeV$^2$) with a  gluon density obtained using starting scale  $\mu_0^2 =1.9$~GeV$^2$ (all other settings are the same as in Set~2) at a scale of $Q^2 =3$~GeV$^2$.  The fit with a starting scale $\mu_0^2 =1.9$~GeV$^2$ gives a $\chi^2=1402.4$ compared to  $\chi^2=1369.8$ when using $\mu_0^2 =1.4$~GeV$^2$. The uncertainties for the new fit include only the uncertainties from experimental sources, the uncertainties for Set~2 are the same as in Fig~\ref{collinear_pdfs_ratio}. Both sets agree within uncertainties. 

\begin{figure}[htb]
\begin{center} 
\includegraphics[width=0.495\textwidth]{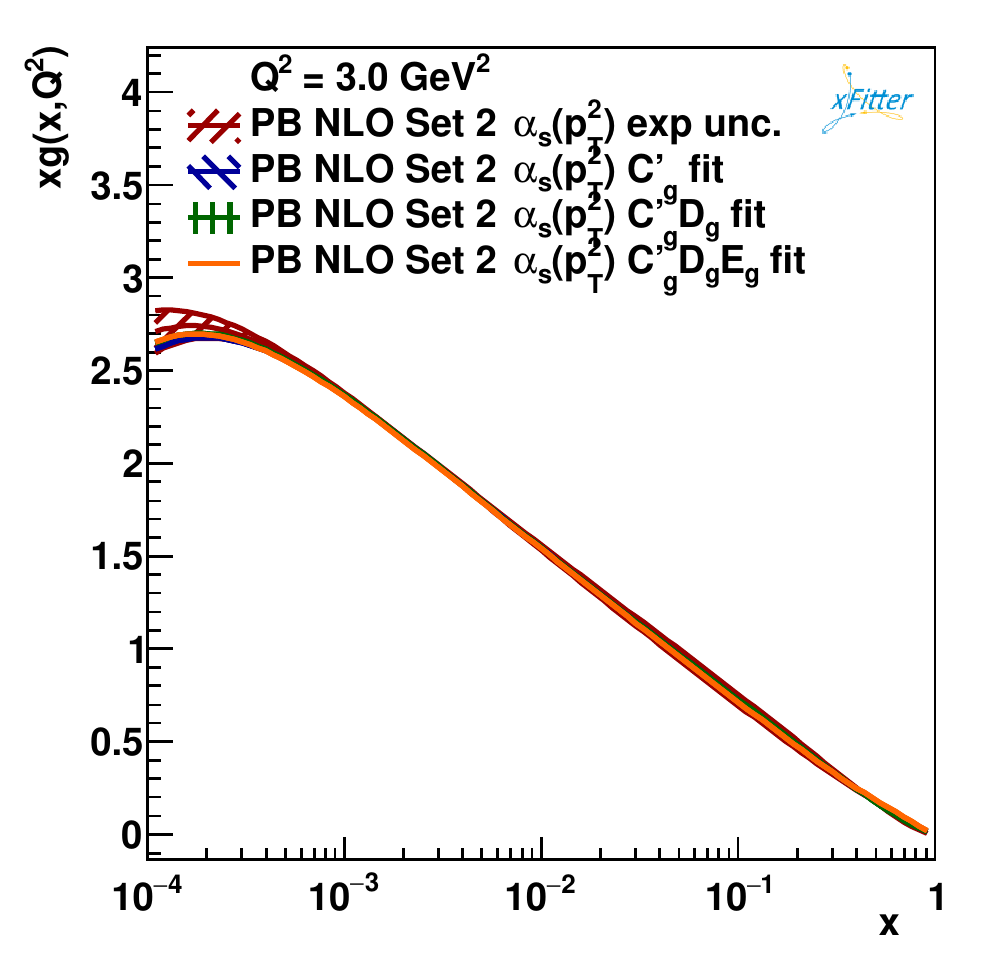} 
\includegraphics[width=0.495\textwidth]{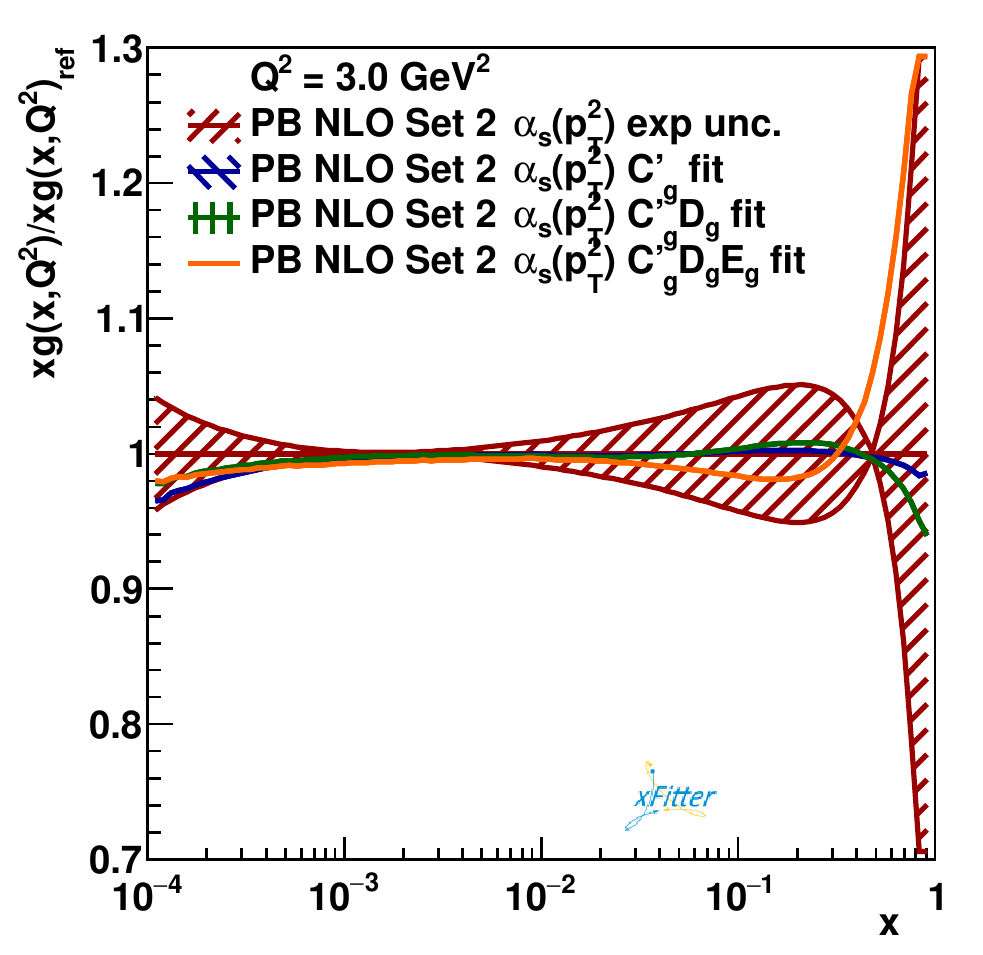} 
\caption{\small Comparison of gluon densities after fit when additional terms in the gluon parametrization are included. 
The uncertainty band of Set~2 corresponds to the uncertainties coming form the experimental sources. }
\label{CgDg}
\end{center}
\end{figure}

A potential bias of the form of the parameterization was checked by extending the 
original parameterization $xg(x) =  A_g x^{B_g} (1-x)^{C_g}  - A'_g x^{B'_g} (1-x)^{C'_g}$ with additional parameters:
\begin{equation}
xg(x) =    A_g x^{B_g} (1-x)^{C_g} (1 + D_g x + E_g x^2)  - A'_g x^{B'_g} (1-x)^{C'_g} . \nonumber
\end{equation}
In Fig.~\ref{CgDg} we show the gluon distribution after fitting $C'_g$ and including the additional factors  $D_g$ and $E_g$ one  after the other. The starting scale is $\mu_0^2  = 1.4$ GeV$^2$ (as for the original fit Set~2). The obtained $\chi^2$  is larger by 1 unit after including additional terms,  the shape of the distribution does not change significantly. The uncertainty band of Set~2 corresponds to the uncertainties coming form the experimental sources, no model or parameterization uncertainty is included. The parton distributions agree within the uncertainties shown, excluding a significant bias from the chosen form of the parametrization. 

\vskip 0.6cm 
\providecommand{\href}[2]{#2}\begingroup\raggedright\endgroup

\end{document}